\newcommand \vers {v2.0}

\documentclass[11pt,preprint]{aastex}

\usepackage{graphicx}
\newcommand{\bracket}[1]{\left\langle#1\right\rangle}

\tightenlines

\newcommand     \beq    {\begin{equation}}
\newcommand     \beqa   {\begin{eqnarray}}
\newcommand     \cm     {\,{\rm cm}}

\newcommand     \eeq    {\end{equation}}
\newcommand     \eeqa   {\end{eqnarray}}

\newcommand     \g      {\,{\rm g}}   
\newcommand     \gtsim  {\gtrsim}                

\newcommand     \ltsim  {\lesssim}               

\newcommand     \aeff   {a_{\rm eff}}
\newcommand     \Rabc    {R_{abc}}
\newcommand     \RKBM   {R_{\rm KBM}}
\newcommand     \Rproj   {R_{\rm proj}}
\newcommand     \ffill   {f}
\newcommand     \fvac    {f_{\rm vac}}     
\newcommand     \poro    {{\cal P}}        
\newcommand     \ndsph    {n_{\rm dip}}    

\countdef\decade=200 \decade=0 \advance\decade by \year
\countdef\hours=201 \hours=0 \advance\hours by \time \divide\hours
by 60 \countdef\mins=202 \mins=0 \advance\mins by \hours
\multiply\mins by 60 \multiply\hours by 100 \countdef\miltime=203
\advance\miltime by -\mins
\newcommand     \todayd{\number\decade.\number\month.\number\day.\number\miltime}
\begin{document}

\title{Modeling Porous Dust Grains with Ballistic Aggregates I: Geometry and Optical
Properties}

\shorttitle{Aggregates as Porous Dust Grains}

\shortauthors{Shen, Draine \& Johnson}

\author{Yue Shen, B. T. Draine, and Eric T. Johnson}

\affil{Princeton University Observatory, Princeton, NJ 08544, USA}

\begin{abstract}
We investigate the scattering and absorption of light by random
ballistic aggregates of spherical monomers. We present general
measures for the size, shape, and
porosity of an irregular particle. Three different
classes of ballistic aggregates are considered, with different
degrees of porosity. Scattering and absorption cross sections are
calculated, using the discrete dipole approximation (DDA), for
grains of three compositions ($50\%$ silicate and $50\%$ graphite;
$50\%$ silicate and $50\%$ amorphous carbon; and $100\%$ silicate,
where percentages are by volume), for
%
wavelengths
from $0.1\ \mu$m to $4\,\mu$m. For fixed amount of solid material,
increased porosity increases the extinction at short wavelengths,
but decreases the extinction at wavelengths long compared to the
overall aggregate size. Scattering and absorption cross sections
are insensitive to monomer size as long as the constituent
monomers are small compared with the incident wavelength. We
compare our accurate DDA results with two other approximations:
the analytical multi-layer sphere (MLS) model and effective medium
theory (EMT). For high porosity and/or absorptive materials, the
MLS model does not provide a good approximation for scattering and
absorption by ballistic aggregates. The EMT method provides a much
better approximation than the MLS model for these aggregates, with
a typical difference $\lesssim 20\%$ in extinction and scattering
cross sections compared with DDA results, for all types,
compositions and wavelengths probed in this study.
\end{abstract}

\section{Introduction}

The appearance of star-forming galaxies is strongly affected by
interstellar dust.  Starlight is absorbed and scattered by dust
particles, and the absorbed energy is predominantly reradiated at
infrared wavelengths. The optical properties of the dust must be
characterized in order to determine the spectra and spatial
distribution of the stars (i.e., to ``correct'' for reddening and
extinction), and to interpret the observed infrared emission.
Characterizing the dust is also important for more fundamental
reasons: dust grains play an important role in the thermodynamics
and chemistry of the interstellar medium (e.g., heating from
photoelectric emission, and grain catalysis of H$_2$), in
interstellar gas dynamics (e.g., radiation pressure on dust
grains), and in the formation of stars, planets, and
planetesimals.

Many interplanetary dust particles (IDPs) are irregular, fluffy
aggregates, sometimes described as having a ``fractal''
appearance. Direct evidence comes from such IDPs collected by
high-flying aircraft
\citep{Brownlee_1985,Warren+Barrett+Dodson+etal_1994}. Laboratory
and microgravity experiments of dust particle interactions, which
may mimic the conditions in the early solar system, also suggest
that particles form fractal assemblies through ballistic
aggregation \citep{Wurm+Blum_1998,
       Blum+Wurm_2000,
       Krause+Blum_2004}.
Similar aggregation processes may well occur in interstellar
environments, where small dust grains coagulate in dense molecular
clouds \citep{Dorschner+Henning_1995}. A number of authors have
proposed that interstellar grains consist primarily of such
aggregate structures, with a random, porous geometry
\citep[e.g.,][]{Mathis+Whiffen_1989}.

Aggregates are irregular, porous, particle assemblies, possibly
incorporating multiple materials. Early attempts to estimate the
optical properties of such structures used a combination of Mie
theory and effective medium theory (EMT), where an aggregate is
approximated by a homogeneous sphere with an ``effective''
refractive index intended to allow for the effects of porosity
\citep[e.g.,][]{Mathis_1996,
                Li+Greenberg_1998}.
Alternatively, \citet{Voshchinnikov+Mathis_1999} proposed a
multi-layer sphere (MLS) model to account for porous, composite
dust grains. The advantage of this MLS model is that, just as for
homogeneous spheres, the solution is exact and can be obtained via
a fast algorithm \citep{Wu+Guo+Ren+etal_1997}. The limitations
are: the grains are assumed to be spherical, and all compositions
(including vacuum) are assumed to be well mixed inside the sphere.
Comparisons of the MLS with EMT-Mie theory are done by
\citet{Voshchinnikov+Ilin+Henning_2005,
       Voshchinnikov+Ilin+Henning+Dubkova_2006},
suggesting that the MLS model is accurate for
spherical,
well-mixed
dust grains, for certain compositions and
porosities. So far there has been no study of whether or not the
MLS prescription is also suitable for irregular aggregates.

Solving Maxwell's equations becomes increasingly challenging when
the geometry departs from spheres. Several methods have been
developed to compute light scattering by non-spherical particles.
The ``extended boundary condition method'' (EBCM) introduced by
\citet{Waterman_1971}, and developed by
\citet{Mishchenko+Travis_1994} and
\citet{Wielaard+Mishchenko+Macke+Carlson_1997} can be applied to
solid targets with relatively smooth surfaces, and exact series
expansions have been developed for spheroids
\citep{Asano+Yamamoto_1975,Asano+Sato_1980,Voshchinnikov+Farafonov_1993,
Farafonov+Voshchinnikov+Somsikov_1996}, but these techniques
cannot be applied to the complex geometries of interest here. The
special case of clusters of spheres can also be treated by
superposition of vector spherical harmonics
\citep{Mackowski_1991,Xu_1997}, sometimes referred to as the
generalized multisphere Mie (GMM) solution. If the complete
T-matrix can be found for the individual particles, the T-matrix
for a cluster can, in principle, be found using the superposition
T-matrix method \citep[TMM, e.g.,][]{Mackowski+Mishchenko_1996}.
However, these techniques, although formally exact, can prove very
computationally-demanding when applied to spheres that are
numerous and in close proximity (e.g., in contact), particularly
when the refractive index differs appreciably from vacuum.

The discrete dipole approximation (DDA) method
\citep[e.g.,][]{Purcell+Pennypacker_1973,
                 Draine+Flatau_1994}
is the most flexible among all the methods, although substantial
computational resources may be needed in order to achieve the desired
accuracy. Using either GMM, TMM or DDA, the optical properties of
different kinds of aggregates have been investigated extensively
during the past decade \citep[e.g.,][]{West_1991,
                Lumme+Rahola_1994,
                Petrova+Jockers+Kiselev_2000,
                Kimura+Kolokolova+Mann_2006,
                Bertini+Thomas+Barbieri_2007}; most of these investigations are
applied to cometary dust.

There is increasing interest in the possibility that interstellar
and circumstellar grains may be porous, random aggregates. To
confront this hypothesis with observational data, it is necessary
to calculate the optical properties of such aggregates. The
calculations must be carried out for a wide range of wavelengths -
from infrared to vacuum ultraviolet - and for aggregates with
potentially realistic geometry and composition. Massive
computations are required to attain these goals.

In this paper, we use the DDA to calculate scattering and
absorption for a large sample of aggregates with different
compositions, sizes, and porosities, over a wide range of
wavelength. We explore the dependence of various optical
properties on wavelength, grain size, aggregate geometry, as well
as compositions, and we investigate the applicability of
multi-layer sphere calculations and the EMT-Mie model to these
random aggregates.

Characterization of the size, gross shape,
and porosity of an irregular particle is
discussed in \S \ref{sec:geom}. \S \ref{sec:model} describes
procedures for generating random aggregates (``BAM1'' and
``BAM2'') that are less porous than the standard ballistic
aggregates, and \S \ref{sec:DDA} discusses the applicability of
the DDA method to these aggregates.

Extinction and absorption behaviors are presented in \S
\ref{sec:extinction},
with comparison to the MLS model in \S \ref{subsec:mls},
and comparison to the EMT-Mie model in \S \ref{subsec:emt}.
Our results are summarized and discussed in
\S\ref{sec:discussion}. Applications to circumstellar debris disks
and cometary dust will be presented in a companion paper
\citep[][hereafter Paper II]{Shen+Draine+Johnson_2007b}.

\section{Geometric Properties of Irregular Particles}\label{sec:geom}

In general, we define the ``effective radius'' (or ``volume
equivalent radius'') of a general 3-dimensional structure
to be the radius of a sphere with volume equal to the volume $V$ of solid
material in the structure:
\beq \label{eq:aeff} \aeff \equiv
\left(\frac{3V}{4\pi}\right)^{1/3}\ .
\eeq

There are several possible ways to quantify the
``porosity'' or openness of the structure, depending on how to
measure its apparent size. Different definitions of the apparent
size for a fluffy structure have been used, i.e., based on the
radius of gyration \citep[e.g.,][]{Kozasa+Blum+Mukai_1992}
or using the geometric
cross section \citep[e.g.,][]{Ossenkopf_1993}.

Here we propose a
simple but general way to quantify the size, shape and porosity of
a finite structure with arbitrary geometry.
Let the ``density''
$\rho_0=1$ at points within solid material, and $\rho_0=0$
otherwise.\footnote{%
   We are discussing purely geometric properties of the target, which
   are independent of the actual material density.
   The function $\rho_0$ is here introduced to allow
   use of familiar concepts like total mass and moment-of-inertia tensor,
   but should be understood to indicate ``occupation'' if $\rho_0=1$,
   and vacuum if $\rho_0=0$.}
Let ${\bf I}$ be
the moment of inertia tensor of the structure, with eigenvalues
$I_1\geq I_2 \geq I_3$. It is convenient to define dimensionless
quantities \beq \alpha_i \equiv \frac{I_i}{0.4 M\aeff^2} \eeq
where $M\equiv \rho_0 V$.
%
For a target that is a solid sphere, $\alpha_i=1$.

We characterize the size and shape of an irregular
structure by considering an ``equivalent ellipsoid'' of uniform
density $f\rho_0$ and semi-major axes $a\leq b\leq c$. The
quantities $(f,a,b,c)$ are uniquely determined by requiring that
the mass $M$ and principal components of the moment of inertia
$I_i$ of the target and its equivalent ellipsoid be identical,
such that:
\beqa
\frac{4\pi}{3}\rho_0 \aeff^3 &=& \ffill\rho_0 \frac{4\pi}{3}abc
\\
\frac{8\pi}{15}\rho_0 \aeff^5 \alpha_1 &=& \ffill\rho_0
\frac{4\pi}{15}abc(b^2+c^2)
\\
\frac{8\pi}{15}\rho_0 \aeff^5 \alpha_2 &=& \ffill\rho_0
\frac{4\pi}{15}abc(a^2+c^2)
\\
\frac{8\pi}{15}\rho_0 \aeff^5 \alpha_1 &=& \ffill\rho_0
\frac{4\pi}{15}abc(a^2+b^2)
\eeqa
with solutions
\beqa \label{eq:a}
a &=&
(\alpha_2+\alpha_3-\alpha_1)^{1/2}\aeff
\\ \label{eq:b}
b &=& (\alpha_3+\alpha_1-\alpha_2)^{1/2}\aeff
\\ \label{eq:c}
c &=& (\alpha_1+\alpha_2-\alpha_3)^{1/2}\aeff
\\ \label{eq:fill factor}
\ffill &=& \left[(\alpha_2+\alpha_3-\alpha_1)
      (\alpha_3+\alpha_1-\alpha_2)
      (\alpha_1+\alpha_2-\alpha_3)\right]^{-1/2} ~~~.
\eeqa
$\ffill$ from eq.\ (\ref{eq:fill factor})
is then the volume filling factor for the
equivalent ellipsoid.
We define a characteristic size
\beq \label{eq:Rabc}
\Rabc\equiv(abc)^{1/3} =
\left[(\alpha_2+\alpha_3-\alpha_1)
      (\alpha_3+\alpha_1-\alpha_2)
      (\alpha_1+\alpha_2-\alpha_3)\right]^{1/6}\aeff
\eeq
which is simply the radius of a sphere with
a volume equal to the
equivalent ellipsoid
defined above;
the filling factor is just $\ffill = (\aeff/R_{abc})^3$.
We define the ``porosity'' to be
\beq \label{eq:poro}
\poro
\equiv 1 - \ffill = 1 - \left(\aeff/R_{abc}\right)^3~~~.
\eeq
\citet{Kozasa+Blum+Mukai_1992}
characterized the overall size of the structure by a
radius proportional
to the radius of gyration:
\beq \label{eq:RKBM}
\RKBM \equiv
\left[\frac{\alpha_1+\alpha_2+\alpha_3}{3}\right]^{1/2}\aeff ~~~.
\eeq
For a uniform density target, $\RKBM = (5/3)^{1/2}R_{\rm gyr}$,
where $R_{\rm gyr}$ is the radius of gyration.
\citet{Kozasa+Blum+Mukai_1992} proposed that the porosity be based on
the ratio between $\RKBM$ and $\aeff$:
\beq \label{eq:poroKBM}
P_{\rm KBM} \equiv 1 - \left(\aeff/\RKBM\right)^3
\eeq
Finally, one can also define a radius
\beq \label{eq:Rproj}
\Rproj\equiv\left[\langle C_{\rm proj}\rangle/\pi\right]^{1/2} ~,
\eeq
where $\langle C_{\rm proj}\rangle$ is the orientation-averaged
projected area of the target.  For irregular targets, $\Rproj$ is
more difficult to compute than either $R_{abc}$ or $\RKBM$.
These three radii -- $\Rabc$, $\RKBM$, and $\Rproj$ --
will be compared below for the ballistic aggregates in this study.

The gross shape of the target can be characterized by the ratios
$c/b$ and $b/a$, where $a,b,c$ are given by eq.\ (\ref{eq:a}-\ref{eq:c}).
We will refer to agglomerates as
oblate if $c/b<b/a$, and prolate if $c/b > b/a$.

It is convenient to define dimensionless efficiency factors $Q$
for absorption, scattering and extinction:
\beq \label{eq:Qdef}
Q(\lambda)\equiv
\frac{C(\lambda)}{\pi \aeff^2} ~~~,
\eeq
where $C(\lambda)$ is the
cross-section for absorption, scattering, or extinction, and
$\aeff$ is the solid-volume-equivalent radius, defined by eq.\ (\ref{eq:aeff}).

\section{Geometric Properties of Ballistic Aggregates}\label{sec:model}

We use 3 simple algorithms for generating irregular porous
structures with varying degrees of porosity.
The first is standard ``ballistic agglomeration'' (BA), sometimes
referred to as ``ballistic particle-cluster agglomeration'' (BPCA),
previously discussed by many authors
\citep[e.g.,][]{%
   West_1991,
   Kozasa+Blum+Mukai_1992,
   Kozasa+Blum+Okamoto+Mukai_1993,
   Ossenkopf_1993,
   Kimura+Kolokolova+Mann_2006,
   Bertini+Thomas+Barbieri_2007}.
We also introduce two new prescriptions for agglomeration -- BAM1
and BAM2 -- that produce random structures that are less
``fluffy'' than those produced by BA (see \S\ref{subsec:targets}).

Random aggregates can be characterized by the Haussdorf dimension
$D$ (sometimes referred to as the fractal dimensionality), with
mass $M \propto R^D$ as $R\rightarrow\infty$ where $R$ is some
characteristic size of the agglomerate. The filling factor $f
\propto M/R^3 \propto M^{1-3/D}$. All three agglomeration
algorithms employed here (BA, BAM1, BAM2) are thought to have
Haussdorf dimension $D\approx3$, with the filling factor
$f\rightarrow$~constant as $M\rightarrow\infty$.

Other processes have also been proposed for generating random
agglomerates, including ballistic cluster-cluster aggregation''
(BCCA) and diffusion-limited aggregation (DLA). BCCA clusters
obtained by random aggregation of clusters with equal-mass
clusters are very ``fluffy'', with $D\approx2.25$
\citep{Kozasa+Blum+Okamoto+Mukai_1993}
 and
filling factor $f\propto M^{-0.33}$.
Clusters produced by diffusion-limited aggregation have a dendritic
appearance, with $D\approx2.5$ \citep{Witten+Cates_1986},
and $f\propto M^{-0.2}$.  We do not consider either BCCA clusters or DLA
clusters here, because their fragile geometries seem unlikely to
be representative of circumstellar or interstellar grains, which are
subject to occasional grain-grain collisions.

We stress that we do not claim that any of our algorithms (BA, BAM1, or BAM2)
provides a
realistic representation of the actual processes responsible for the
growth of circumstellar or interstellar grains -- they are merely
convenient procedures for generating irregular targets that may
bear some geometric resemblance to real circumstellar or interstellar grains.

\subsection{Target Generation}\label{subsec:targets}

We construct clusters (``targets'') by random ballistic
agglomeration. Each cluster is composed of a certain number of
spheres (or ``monomers'') of a single radius $a_0$. The
agglomeration process begins with one monomer ($j=1$). A cluster
is built up by sequential arrival of additional monomers ($j=2,
..., N$) on random rectilinear trajectories.  We consider three
different classes of clusters, distinguished by aggregation rules.

\begin{figure*}
  \centering
    \includegraphics[angle=0,width=\textwidth]{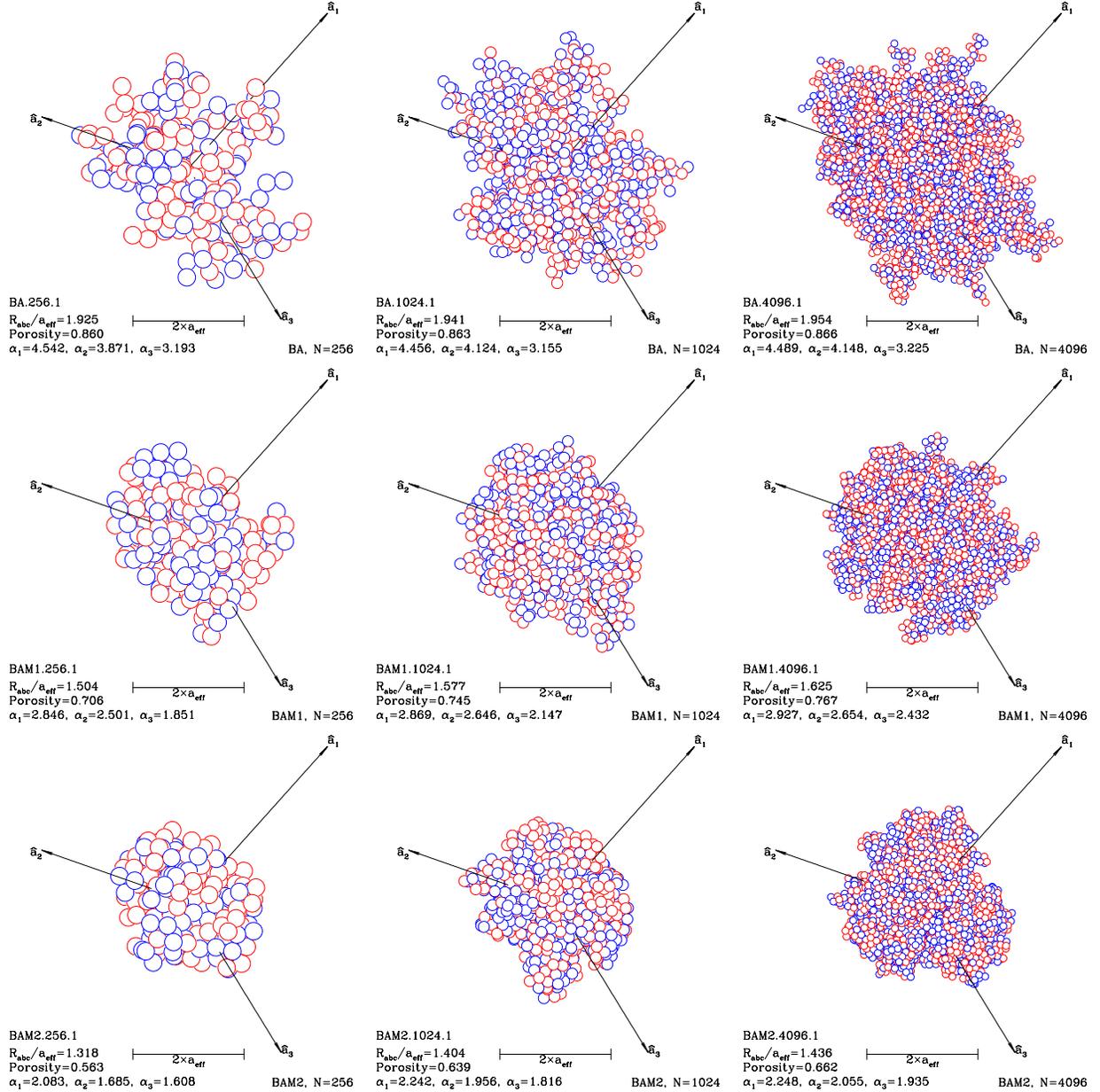}
    \caption{\label{fig:BA_BAM1_BAM2}\footnotesize
    From {\em left} to {\em right}: N=256, 1024, and 4096 clusters.
    From {\em top} to {\em bottom}: BA, BAM1, and BAM2 aggregation rules.
    Axes $\hat{\bf a}_1$, $\hat{\bf a}_2$, $\hat{\bf a}_3$ are the principal
    axes with the largest, intermediate, and smallest moment of inertia.
    For each cluster we give the characteristic radius $R_{abc}$,
    the porosity $\poro$, and the dimensionless moment-of-inertia eigenvalues
    $\alpha_i$.}
\end{figure*}

BA (``ballistic agglomeration'') clusters are produced by
requiring arriving monomers to ``stick'' at the point where they
first contact the preexisting aggregate. This is a
well-established procedure \citep[e.g.,][]{%
   West_1991,
   Kozasa+Blum+Mukai_1992,
   Kozasa+Blum+Okamoto+Mukai_1993,
   Ossenkopf_1993,Kimura+Kolokolova+Mann_2006,
   Bertini+Thomas+Barbieri_2007},
also known as ``Ballistic Particle-Cluster Aggregation'' (BPCA).
Clusters formed in this way have fractal dimension $\approx3$, but
high porosity. Figure \ref{fig:BA_BAM1_BAM2} (top row) shows
examples of BA clusters with $N=256,\, 1024,\, 4096$ monomers.

BAM1 (``ballistic agglomeration with one migration'') clusters are
produced by requiring arriving monomers $j\geq 3$, after making
first contact with a monomer $k<j$, to ``migrate'' to make contact
with another monomer, by rolling or sliding over the
first-contacted monomer, along the shortest possible trajectory.
If there is more than one candidate for this second contact, the
nearer is chosen. BAM1 clusters with $N\geq3$ have every monomer
in contact with at least two other monomers; for $N\geq4$ some of
the monomers are in contact with three or more other monomers.
Figure \ref{fig:BA_BAM1_BAM2} (middle row) shows BAM1 clusters
with $N=256,\, 1024,\, 4096$ monomers.

BAM2 clusters are constructed as follows: monomers $j=2$ and 3 are
added randomly just as for the BAM1 clusters. Monomers $j\geq 4$
arrive on random rectilinear trajectories; after first contact
they make {\it two} migrations.  The first migration is the same
as for constructing BAM1 clusters: ``rolling'' along the shortest
possible trajectory to make a second contact. This is followed by
a second migration, now rolling over both the first and second
sphere contacted to contact a third neighbor, again choosing the
shortest trajectory if there is more than one candidate. BAM2
clusters with $N\geq4$ have every monomer in contact with at least
three neighbors, with some in contact with four or more neighbors.
Figure \ref{fig:BA_BAM1_BAM2} (bottom row) shows examples of BAM2
clusters with $N=256,\, 1024,\, 4096$ monomers.

For clusters consisting of $N$ spherical monomers of radius $a_0$
we have, \beq \aeff = N^{1/3} a_0\ . \eeq

We generated clusters using the BA, BAM1, and BAM2 algorithms, for
a large number of different seeds for the random number generator.
A library of samples is available\footnote{
\url{http://www.astro.princeton.edu/$\sim$draine/agglom.html}}
for $N=2^3,2^4,...,2^{16}$. Below we report the statistical properties
of such clusters.

\subsection{Size and Porosity of Ballistic Aggregates}

The clusters formed following the above procedure are generally
irregular and ``porous'', with the porosity decreasing from BA to
BAM1 to BAM2, as is apparent from comparison of the three rows in
Fig. \ref{fig:BA_BAM1_BAM2}.

\begin{deluxetable}{r c c c c c c c c c}
\tabletypesize{\footnotesize} \rotate \tablewidth{0pt}
\tablecolumns{10} \tablecaption{\label{tab:cluster sizes}
            Sizes of BA, BAM1, and BAM2 Clusters}
\tablehead{
  \colhead{}
  & \multicolumn{3}{c}{Characteristic Radius $\Rabc/\aeff\,^{a}$
             from eq.\ (\ref{eq:Rabc})}
  & \multicolumn{3}{c}{Radius $\RKBM\,^{a}$
             from eq.\ (\ref{eq:RKBM})}
  & \multicolumn{3}{c}{Projected Area Radius $\Rproj\,^{b}$}}
\startdata
$N$  &  BA & BAM1 & BAM2 &  BA & BAM1 & BAM2 & BA & BAM1 & BAM2 \\
8      &  $1.4300\pm0.0732$ & $1.2435\pm0.0287$ & $1.1569\pm0.0136$
       &  $1.7049\pm0.1683$ & $1.3786\pm0.0650$ & $1.2133\pm0.0230$
       &  $1.2569\pm0.0163$ & $1.1979\pm0.0116$ & $1.1458\pm0.0057$
\\
16     &  $1.5885\pm0.0868$ & $1.3304\pm0.0367$ & $1.1860\pm0.0167$
       &  $1.8209\pm0.1671$ & $1.4611\pm0.0828$ & $1.2526\pm0.0453$
       &  $1.3514\pm0.0197$ & $1.2670\pm0.0158$ & $1.1861\pm0.0106$
\\
32     &  $1.7060\pm0.0884$ & $1.4079\pm0.0440$ & $1.2212\pm0.0228$
       &  $1.8906\pm0.1486$ & $1.5242\pm0.0838$ & $1.2910\pm0.0521$
       &  $1.4469\pm0.0219$ & $1.3362\pm0.0183$ & $1.2288\pm0.0138$
\\
64     &  $1.7861\pm0.0794$ & $1.4733\pm0.0454$ & $1.2614\pm0.0271$
       &  $1.9262\pm0.1227$ & $1.5684\pm0.0751$ & $1.3258\pm0.0501$
       &  $1.5405\pm0.0227$ & $1.4036\pm0.0194$ & $1.2727\pm0.0155$
\\
128    &  $1.8371\pm0.0668$ & $1.5262\pm0.0427$ & $1.2998\pm0.0284$
       &  $1.9403\pm0.0966$ & $1.5991\pm0.0628$ & $1.3530\pm0.0450$
       &  $1.6299\pm0.0225$ & $1.4681\pm0.0193$ & $1.3146\pm0.0165$
\\
256    &  $1.8705\pm0.0535$ & $1.5650\pm0.0356$ & $1.3340\pm0.0271$
       &  $1.9464\pm0.0745$ & $1.6190\pm0.0498$ & $1.3759\pm0.0381$
       &  $1.7138\pm0.0216$ & $1.5277\pm0.0182$ & $1.3552\pm0.0165$
\\
512    &  $1.8914\pm0.0418$ & $1.5932\pm0.0296$ & $1.3633\pm0.0245$
       &  $1.9470\pm0.0565$ & $1.6325\pm0.0396$ & $1.3949\pm0.0323$
       &  $1.7894\pm0.0199$ & $1.5814\pm0.0171$ & $1.3940\pm0.0161$
\\
1024   &  $1.9052\pm0.0329$ & $1.6157\pm0.0243$ & $1.3899\pm0.0202$
       &  $1.9454\pm0.0429$ & $1.6428\pm0.0297$ & $1.4130\pm0.0254$
       &  $1.8556\pm0.0183$ & $1.6289\pm0.0158$ & $1.4315\pm0.0144$
\\
2048   &  $1.9177\pm0.0248$ & $1.6343\pm0.0183$ & $1.4114\pm0.0171$
       &  $1.9467\pm0.0322$ & $1.6535\pm0.0225$ & $1.4274\pm0.0207$
       &  $1.9131\pm0.0160$ & $1.6708\pm0.0141$ & $1.4647\pm0.0130$
\\
4096   &  $1.9286\pm0.0198$ & $1.6500\pm0.0151$ & $1.4300\pm0.0135$
       &  $1.9493\pm0.0250$ & $1.6630\pm0.0175$ & $1.4409\pm0.0159$
       &  $1.9612\pm0.0148$ & $1.7060\pm0.0126$ & $1.4934\pm0.0110$
\\
8192   &  $1.9384\pm0.0151$ & $1.6631\pm0.0115$ & $1.4462\pm0.0114$
       &  $1.9532\pm0.0174$ & $1.6718\pm0.0130$ & $1.4533\pm0.0122$
       &  $2.0002\pm0.0130$ & $1.7344\pm0.0101$ & $1.5177\pm0.0095$
\\
16384  &  $1.9508\pm0.0121$ & $1.6758\pm0.0097$ & $1.4588\pm0.0082$
       &  $1.9606\pm0.0129$ & $1.6828\pm0.0113$ & $1.4640\pm0.0088$
       &  $2.0334\pm0.0110$ & $1.7574\pm0.0093$ & $1.5363\pm0.0068$
\\
32768  &  $1.9620\pm0.0089$ & $1.6875\pm0.0073$ & $1.4714\pm0.0063$
       &  $1.9691\pm0.0091$ & $1.6926\pm0.0089$ & $1.4751\pm0.0066$
       &  $2.0586\pm0.0079$ & $1.7762\pm0.0073$ & $1.5531\pm0.0065$
\\
65536  &  $1.9771\pm0.0076$ & $1.6996\pm0.0056$ & $1.4835\pm0.0036$
       &  $1.9811\pm0.0089$ & $1.7032\pm0.0067$ & $1.4862\pm0.0040$
       &  $2.0795\pm0.0086$ & $1.7924\pm0.0063$ & $1.5664\pm0.0041$
\\
\hline
\multicolumn{10}{l}{$^a$ Mean and $\pm1\sigma$ realization-to-realization
variation in $\Rabc$ from eq.\ (\ref{eq:Rabc}).}\cr
\multicolumn{10}{l}{$^a$ Mean and $\pm1\sigma$ realization-to-realization
variation in $\RKBM$ from eq.\ (\ref{eq:RKBM}).}\cr
\multicolumn{10}{l}{$^b$ Mean and $\pm1\sigma$ realization-to-realization
variation in $\Rproj$ from eq. (\ref{eq:Rproj}).}
\enddata
\end{deluxetable}
\begin{deluxetable}{r c c c c c c c c c}
\tabletypesize{\footnotesize}
\tablewidth{0pt}
\tablecolumns{7}
\tablecaption{\label{tab:cluster poros}
            Porosities of BA, BAM1, and BAM2 Clusters}
\tablehead{
  \colhead{}
  & \multicolumn{3}{c}{Porosity $\poro\,^{a}$ from eq.\ (\ref{eq:poro})} &
    \multicolumn{3}{c}{Porosity $P_{\rm KBM}\,^{b}$}}
\startdata
$N$  & BA & BAM1 & BAM2 & BA & BAM1 & BAM2 \\
8      &  $0.6580\pm0.0525$ & $0.4799\pm0.0360$ & $0.3541\pm0.0228$
       &  $0.7982\pm0.0598$ & $0.6183\pm0.0540$ & $0.4402\pm0.0319$
\\
16     &  $0.7505\pm0.0409$ & $0.5753\pm0.0351$ & $0.4006\pm0.0253$
       &  $0.8265\pm0.0443$ & $0.6736\pm0.0519$ & $0.4874\pm0.0522$ \\

32     &  $0.7986\pm0.0313$ & $0.6417\pm0.0336$ & $0.4509\pm0.0307$
       &  $0.8468\pm0.0338$ & $0.7127\pm0.0444$ & $0.5309\pm0.0532$ \\

64     &  $0.8245\pm0.0234$ & $0.6873\pm0.0289$ & $0.5018\pm0.0321$
       &  $0.8568\pm0.0259$ & $0.7374\pm0.0359$ & $0.5674\pm0.0467$ \\

128    &  $0.8387\pm0.0176$ & $0.7187\pm0.0236$ & $0.5446\pm0.0299$
       &  $0.8611\pm0.0200$ & $0.7533\pm0.0280$ & $0.5937\pm0.0390$ \\

256    &  $0.8472\pm0.0131$ & $0.7391\pm0.0178$ & $0.5788\pm0.0257$
       &  $0.8632\pm0.0152$ & $0.7630\pm0.0213$ & $0.6144\pm0.0312$ \\

512    &  $0.8522\pm0.0098$ & $0.7527\pm0.0138$ & $0.6053\pm0.0213$
       &  $0.8638\pm0.0115$ & $0.7694\pm0.0164$ & $0.6304\pm0.0250$ \\

1024   &  $0.8554\pm0.0075$ & $0.7629\pm0.0107$ & $0.6276\pm0.0162$
       &  $0.8638\pm0.0089$ & $0.7740\pm0.0121$ & $0.6449\pm0.0189$ \\

2048   &  $0.8582\pm0.0055$ & $0.7709\pm0.0077$ & $0.6443\pm0.0129$
       &  $0.8642\pm0.0066$ & $0.7786\pm0.0090$ & $0.6558\pm0.0149$ \\

4096   &  $0.8606\pm0.0043$ & $0.7774\pm0.0061$ & $0.6580\pm0.0097$
       &  $0.8649\pm0.0051$ & $0.7824\pm0.0068$ & $0.6655\pm0.0110$ \\

8192   &  $0.8627\pm0.0032$ & $0.7826\pm0.0045$ & $0.6694\pm0.0078$
       &  $0.8657\pm0.0036$ & $0.7859\pm0.0050$ & $0.6470\pm0.0082$ \\

16384  &  $0.8653\pm0.0025$ & $0.7875\pm0.0037$ & $0.6779\pm0.0054$
       &  $0.8673\pm0.0026$ & $0.7901\pm0.0042$ & $0.6812\pm0.0057$ \\

32768  &  $0.8676\pm0.0018$ & $0.7919\pm0.0027$ & $0.6861\pm0.0040$
       &  $0.8690\pm0.0018$ & $0.7937\pm0.0032$ & $0.6884\pm0.0042$ \\

65536  &  $0.8706\pm0.0015$ & $0.7963\pm0.0020$ & $0.6937\pm0.0022$
       &  $0.8714\pm0.0017$ & $0.7976\pm0.0024$ & $0.6953\pm0.0025$ \\
\hline
\multicolumn{7}{l}{$^a$ Mean and $\pm1\sigma$ realization-to-realization
variation in porosity $\poro$ from eq.\ (\ref{eq:poro}).}\cr
\multicolumn{7}{l}{$^b$ Mean and $\pm1\sigma$ realization-to-realization
variation in porosity $P_{\rm KBM}$ from eq. (\ref{eq:poroKBM}.)}
\enddata
\end{deluxetable}

For a given $N$ and aggregation rule, the characteristic size $\Rabc$
will vary from realization to realization because the aggregation
process is random.
Figure \ref{fig:frchar} shows the expectation value $\langle\Rabc\rangle$
based on
many random realizations, for $N=2^3,2^4,...,2^{16}$; the realization-to-realization variations are also shown.
The BA clusters have $\langle\Rabc\rangle\rightarrow 2.0\aeff$
as $N\rightarrow\infty$,
the BAM1 clusters have $\langle\Rabc\rangle\rightarrow 1.7\aeff$, and the BAM2
clusters have $\langle\Rabc\rangle\rightarrow 1.5\aeff$.
Note, however, that even for $N>10^4$, the $\langle\Rabc/\aeff\rangle$
continues to
increase as $N$ increases from $2^{14}$ to $2^{15}$ to $2^{16}$, so the
limiting values for $N\rightarrow\infty$ are uncertain.
Figure \ref{fig:frchar} also shows $\langle\RKBM\rangle$, the characteristic
size given by eq.\ (\ref{eq:RKBM}).
For $N > 10^4$, $\RKBM\approx\Rabc$.  However, for smaller $N$ $\RKBM$ is
noticeably larger than $\Rabc$.  In addition,
the
realization-to-realization
variation in $\RKBM$ is appreciably larger than
for $\Rabc$.  We consider $\Rabc$ to be the better way to characterize
the effective size of a random structure.

For each of our clusters, we have estimated the
orientation-averaged projected area, and from this the radius
$\Rproj$.  These computations are time-consuming; the computation
is carried out by enclosing the target within a sphere of radius
$R_c$, choosing $N_r$ points at random on the sphere, for each
such point choosing a random inward direction (drawn from a
distribution corresponding to isotropic incidence on the sphere),
and determining whether that ray does or does not intersect the
target structure. If the number of intersecting rays is $N_x$,
then $\Rproj=\sqrt{N_x/N_r}R_c$. We typically employ $N_r=10^6$ in
order to achieve accuracy better than 1\% in the determination of
$\Rproj$.  It is striking that, for a given $N$, $\Rproj$ in
Figure \ref{fig:frchar} shows less realization-to-realization
veriation than even $R_{abc}$.

The expectation values $\langle\ffill\rangle$ and
$\langle\poro\rangle=1-\langle\ffill\rangle$
are shown in Figure \ref{fig:ffill}.
For large $N$, the BA clusters have
$\langle\poro\rangle\approx0.87$, whereas the
BAM2 clusters have substantially lower porosity,
$\langle\poro\rangle\approx0.70$.

Validation of a code to generate random clusters is not simple,
but can be done by comparing the statistical properties of the
clusters with those of other Monte Carlo cluster generators.
Statistical properties of the three types of aggregate clusters
are given in Tables \ref{tab:cluster sizes} and \ref{tab:cluster
poros} for cluster sizes ranging from $N=2^3$ to $N=2^{16}$.

The BA clusters have been frequently used in the literature
\citep[e.g.,][]{Kozasa+Blum+Mukai_1992,
                Kozasa+Blum+Okamoto+Mukai_1993,
                Ossenkopf_1993,
                Kimura+Kolokolova+Mann_2006,
                Bertini+Thomas+Barbieri_2007}.
For the same $N$, our calculated
projected areas agree with results for BA
clusters reported by \citet{Kozasa+Blum+Okamoto+Mukai_1993} and
\citet{Nakamura+Kitada+Mukai_1994} as shown in Figure
\ref{fig:frchar} (top panel).
However, as seen in Figure \ref{fig:frchar} (middle panel),
our values of $\RKBM$ (defined by eq.\ \ref{eq:RKBM})
are slightly lower than the results reported by
\citet{Kozasa+Blum+Okamoto+Mukai_1993}.
The reason for this discrepancy
is not known.\footnote{
    T.\ Kozasa has kindly provided several BA clusters from
    \citet{Kozasa+Blum+Okamoto+Mukai_1993}.
    For these clusters, we obtain values of $\Rabc/\aeff$ and
    $\RKBM/\aeff$ that are consistent with the statistics reported in our
    Table \ref{tab:cluster sizes}.}

\subsection{Shape of Ballistic Aggregates}

The axial ratios $c/b$ and $b/a$ give an indication of the overall
shape of each random aggregate.
Figure \ref{fig:axrat} shows $c/b$ and $b/a$ for BA, BAM1, and BAM2 clusters
with $N=2^6$, $2^8$, $2^{10}$, and $2^{12}$ spheres.
For a given $N$, the sequence BA$\rightarrow$BAM1$\rightarrow$BAM2
corresponds to decreasing eccentricity.
Similarly, for a given agglomeration rule (BA, BAM1, or BAM2), clusters
with larger $N$ tend to be more spherical.

Prolate spheroids would have $b/a=1$; oblate spheroids have
$c/b=1$. Figure \ref{fig:axrat} shows that the clusters are
generally triaxial, with a tendency toward prolateness ($c/b >
b/a$).

The ballistic aggregates constructed in this section
are based on very simple algorithms.
More realistic treatments of grain
coagulation processes are possible
\citep[e.g.,][]{Ossenkopf_1993,Ormel+Spaans+Tielens_2007},
but are rather uncertain
given our limited knowledge of the interstellar/circumstellar
environment and grain properties.
The well-defined sequence of BA
to BAM2 clusters described here allows us to systematically
investigate the grain properties as function of porosity.

\section{Application of the Discrete Dipole Approximation}\label{sec:DDA}
\subsection{Composition}


The composition of interstellar grains continues to be uncertain.
The observed strength of the 10 $\micron$ absorption feature,
produced by the Si-O stretching mode in silicates, requires that
$\gtsim$50\% of the total grain volume be amorphous silicate
material. As discussed  below,
this amount of silicate material is also consistent with
the observed depletions of Si, Mg, and Fe from the gas phase.

Attempts to reproduce the observed wavelength-dependent extinction
require more grain material than can be provided by silicates
alone \citep[e.g,][]{Weingartner+Draine_2001a,Zubko+Dwek+Arendt_2004}.
Because H$_2$O ice is not present in the diffuse
interstellar medium, and Mg, Fe, and Si are presumed to be
primarily invested in the silicate material, the only element that
can provide substantial additional solid material is carbon.
Observational evidence for carbonaceous material includes
\citep[see][and references therein]{Draine_2003a}: (1)
strong absorption near $0.22\,\micron$, likely due to
$\pi\rightarrow\pi^*$ electronic excitation in $sp^2$-bonded
carbon (e.g., graphite, or polycyclic aromatic hydrocarbons); (2)
absorption at $3.4\,\micron$ (attributed to the C-H stretching
mode); and (3) observed emission features at 3.3, 6.2, 7.6, 8.6,
11.3, and $12.7\ \micron$ that are attributed to vibrational modes
of
polycyclic
aromatic hydrocarbons (PAHs). The cosmic abundance of
carbon, and the fact that it is moderately depleted from the gas
phase, is consistent with carbonaceous material with a total solid
volume
that is a substantial fraction of the volume of amorphous silicate material.

Assuming total interstellar abundances of the elements relative to H
to be the same as current estimates
of solar abundances, observations of gas-phase abundances in the
interstellar medium allow us to infer the amount of different elements
locked up in grains \citep[see Table 1 of][]{Draine_2008}.
Current estimates indicate that carbon in grains contributes a mass
of about 0.13\% of the total H mass, while solid material containing
Mg, Fe, Si, O (plus small contributions from
Al, Ca, and Ni) amounts to about 0.57\% of the total H mass.
If the carbonaceous material has a density of $\sim2\g\cm^{-3}$,
and the silicate material has a density $\sim3.5\g\cm^{-3}$ then
the overall
silicate/carbon volume ratio is $V_{\rm sil}/V_{\rm car}\approx 2.5$.
This, however, is based on assuming the total C abundance to be only
$245\pm30$~ppm relative to H \citep{Asplund+Grevesse+Sauval+etal_2005}.
However, two recent determinations of $({\rm O/H})_\odot$
\citep{Landi+Feldman+Doschek_2007,Centeno+Socas-Navarro_2008}
obtain values
that are $\sim1.9$ times larger than the solar oxygen abundance
$({\rm O/H})_\odot=457\pm56$~ppm
of \citet{Asplund+Grevesse+Sauval+etal_2004,Asplund+Grevesse+Sauval+etal_2005}.
The solar C abundance might therefore be larger than
Asplund et al.'s value of 245~ppm.
If the total C abundance were to be increased to,
e.g., 350~ppm, the mass of C in dust
would increase by 75\%, and the silicate/carbonaceous volume
ratio would fall to $V_{\rm sil}/V_{\rm car}\approx 1.4$.
The total abundances of Mg, Fe, Si -- and therefore the inferred abundance
of solids with silicate composition -- are of course also uncertain.
We conclude that if grains are of mixed composition, the silicate:carbonaceous
volume fractions could be as large as 70:30, or as low as 60:40.
We will consider extreme examples of composite
grains that are 100\% silicate, as well
as composite grains with 50:50 volume fractions.

While the importance of carbon is undisputed, the specific form
is uncertain. The observed PAH emission features require
$\gtsim$20\% of the solid carbon to be in small free-flying PAH
molecules or clusters. Spectroscopy of the $3.4\ \micron$ feature
indicates that $\sim15\%$ of the
hydrocarbon material is aliphatic (chain-like), and $\sim85\%$
aromatic ($sp^2$-bonded)
\citep{Pendleton+Allamandola_2002}.
Diamond ($sp^3$-bonded carbon) has been
found in meteorites, and may also be present in the interstellar
medium, but there is no direct evidence for interstellar diamond.

Because the predominant form of carbon is uncertain, we consider
two possibilities: crystalline graphite, and amorphous carbon. For
graphite, a highly anisotropic material, we use the dielectric
tensor from \citet{Draine_2003b}.
Each graphite sphere is assumed to be a single crystal, with
the crystal axes for each graphite monomer assigned an independent
random orientation.

``Amorphous carbon'' is not a
well-defined material, and its properties depend on the method of
preparation. We use the dielectric function for amorphous carbon
``AC1'' from \citet{Rouleau+Martin_1991}. For the amorphous
silicate component, we use the ``astrosilicate'' dielectric
function from \citet{Draine_2003b}.

We will consider three compositions for the clusters: (1) 50\%
amorphous silicate and 50\% graphite; (2) 50\% amorphous silicate
and 50\% AC1; and (3) 100\% amorphous silicate. For the
mixed-composition clusters, the monomer compositions are assigned
randomly, and the percentages are by volume.

\subsection{DDA Method and Validity Criteria}\label{sec:DDA_model}

We calculate the absorption and scattering properties of the BA,
BAM1 and BAM2 clusters using
DDSCAT 7.0 \citep{Draine+Flatau_2008b}.
DDSCAT is a code based on the discrete dipole approximation
\citep{Purcell+Pennypacker_1973,Draine+Flatau_1994}, designed to
compute scattering and absorption of electromagnetic waves by
targets with arbitrary geometry and composition, for targets that
are not too large compared to the wavelength $\lambda$.

There are three validity criteria that should be satisfied for the
DDA to provide accurate results:
\begin{enumerate}
\item The interdipole spacing $d$ should be small enough to
resolve the geometric structure of the target.  This is
accomplished provided the number of dipoles per spherical monomer
$\ndsph\gtsim100$.

\item The phase shift corresponding to one dipole spacing should
be small. \beq\label{eq:dda_c1} |m|kd = 0.44|m|
\left(\frac{100}{\ndsph}\right)^{1/3} \left(\frac{a_0}{0.02\
\micron}\right) \left(\frac{0.1\ \micron}{\lambda}\right) \ltsim 1
\eeq where $m$ is the refractive index, $k\equiv 2\pi/\lambda$,
$a_0$ is the monomer sphere radius, and $\ndsph$ is the number of
dipoles per sphere. Thus we see that $\ndsph=100$ allows
condition (\ref{eq:dda_c1}) to be satisfied even at $\lambda=0.1\
\micron$\footnote{At $\lambda=0.1\ \micron$, $|m|=1.84$, 2.41,
2.15, 1.91 for astrosilicate, graphite with $E\perp c$, graphite
with $E\parallel c$, and AC1.} for $a_0=0.02\ \micron$. For
accurate calculations of the scattering phase function, the
validity requirement is somewhat more stringent, $|m|kd\ltsim0.5$.
With $\ndsph=100$ and $a_0=0.02\ \micron$, we have
$|m|kd\ltsim0.5$ for $\lambda \gtsim 0.2\ \micron$.

\item Even when the criterion $|m|kd<0.5$ is satisfied, the
implementation of the DDA used here overestimates absorption in
materials with ${\rm Im}(\epsilon)\gg 1$, where $\epsilon$ is the
dielectric function. Graphite is a conducting material; the
dielectric tensor component for $E\perp c$ (with $c$ being the
normal to the basal plane) becomes large in the infrared
($\epsilon_\perp=11.7+32.6i$ at $\lambda=3.981\,\micron$), and the
accuracy of the DDA suffers.
\end{enumerate}
To assess the computational accuracy, some of the scattering
calculations have been repeated using differing numbers of dipoles
per monomer $n_{\rm dip}$. In the limit $n_{\rm
dip}\rightarrow\infty$, the DDA becomes exact.  For finite $n_{\rm
dip}$, the errors are expected to scale as $n_{\rm dip}^{-1/3}$
(i.e., the errors vary linearly with the interdipole separation
$d$); such behavior has previously been demonstrated by
\citet{Collinge+Draine_2004} and \citet{Yurkin+Maltsev+Hoekstra_2006a}.

Figure \ref{fig:converg} shows the total extinction and scattering
efficiencies computed for one
cluster
($50\%$ silicate
and $50\%$ graphite) with $N=256$ $a_0=0.02\ \micron$ monomers,
and porosity $\poro=0.853$,
for a single orientation, using different numbers of dipoles with
$\ndsph\approx $ 100, 200, 400, 1000. The results show that the
computed cross sections at each wavelength are approximately
linear functions of $\ndsph^{-1/3}$, allowing us to confidently
extrapolate to estimate the exact result at $n_{\rm
dip}^{-1/3}\rightarrow 0$, and thereby to estimate the error that
would result if we were to simply use cross sections calculated
for $\ndsph\approx100$. On average, the fractional errors for the
results with $\ndsph\approx 100$ are only a few percent for
$\lambda\lesssim 0.35\micron$, 4--10\% for
$0.35 < \lambda < 0.8\micron$, and
10--15\% for $0.8 < \lambda < 4\micron$.
Note that for wavelengths $\lambda\gtsim 0.5\
\micron$, the computations for finite $\ndsph$ always {\it
overestimate} the absorption cross section and scattering cross
sections, presumably as a result of failure to accurately resolve
the ``shielding'' produced by the charge layer produced by the
discontinuity in the polarization field at the grain surface.

The fractional errors of
$\sim10\%$
in the infrared are
comparable to the typical variations in $Q_{\rm abs}$ and $Q_{\rm
ext}$ from one random realization to another. Hence using
$\ndsph\approx100$ dipoles per monomer suffices for most of our
investigations\footnote{
   For clusters with $N\le512$,
   we use results obtained with
   $\ndsph\approx 400$ for improved accuracy; for larger clusters, i.e.,
$N=1024,\ 2048,$ etc., we use $\ndsph\approx 100$ due to
computational limits.}, and allows us to explore the parameter
space more efficiently.

For the DDA results presented in the next section, each
calculation is averaged over a few (3 or 5) realizations and 54
random orientations for each realization.
The typical variations in cross sections from
realization to realization for the same type of clusters are of
the order $\lesssim 10\%$; while 54 random orientations are
sufficient to represent the orientational average
\citep[see][]{Johnson+Draine_2008}. The scattering and absorption
are calculated at 32 wavelengths from $0.1$ to $3.981\ \micron$
(with $\Delta\log_{10}\lambda=0.05, \ 0.025,\ 0.05,\ 0.1$ for
$0.1\le\lambda\le 0.1259$,
 $0.1259<\lambda\le 0.3126$, $0.3126<\lambda\le 0.5012$
 and $0.5012<\lambda\le 3.981\ \micron$, respectively).

\section{Absorption, Scattering, and Extinction Cross Sections of Aggregates}
         \label{sec:extinction}

In \S\S \ref{subsec:dda},\ref{subsec:monomer_size}
we fix the effective radius $\aeff=0.127\ \micron$ for each cluster and
explore the differences in optical properties between three
different cluster geometries and three different compositions. The
aggregates have characteristic radius
$\Rabc\approx 0.16-0.24\micron$.
In \S\ref{subsec:monomer_size} we vary the
monomer size to investigate the sensitivity of results to monomer
size and to porosity at fixed $\aeff$.
In \S\ref{subsec:mls} we compare our DDA results with the
analytical multi-layer sphere model
\citep{Voshchinnikov+Mathis_1999} to test the accuracy of the MLS
prescription for estimating the optical properties of ballistic
aggregates. In \S\ref{subsec:emt} we carry out a similar
comparison of our DDA results with ``effective medium theory''.

%
%
\subsection{DDA Results: Dependence on Wavelength and Porosity}
            \label{subsec:dda}

The orientation- and realization-averaged (5 realizations)
extinction cross section, absorption cross section and scattering
cross section, as well as the asymmetry parameter $g\equiv
\langle\cos\theta\rangle$ are plotted in Figure \ref{fig:DDA}, for
the three aggregate types (BA, BAM1, BAM2) and three different
compositions, as functions of wavelength. We have used the results
computed with $n_{\rm dip}\approx 400$, which are within a few percent of the
exact values as inferred from Fig. \ref{fig:converg}.

For wavelength $\lambda\gg
\aeff$, the clusters are in the Rayleigh limit; the extinction is
dominated by absorption, and the asymmetry parameter
$g\equiv\langle\cos\theta\rangle$ is small. For the same volume of
solid material, the 100\% silicate clusters have smaller
absorption (and therefore total extinction) at long wavelength,
because graphite and AC1 are more absorptive than silicate in the
near-infrared. At short wavelength, the behavior of the cross
sections and $g$ is nonmonotonic.
For example, for the
50\%~silicate/50\%~graphite clusters, $Q_{\rm ext}$ has
a dip at $\sim 0.17\,\micron$ and peak near $0.22\,\micron$.

An important question is how the porosity of the dust grain
affects the extinction efficiency. One might expect that increased
porosity would lead to an increase in the overall extinction cross
section per unit solid material, which motivated attempts to try
to use porous dust grains to circumvent the difficulties in
accounting for the observed interstellar extinction without
overconsuming the elements used to build grains. Fig.
\ref{fig:DDA} shows that for short wavelengths, increased porosity
does result in a modest increase in extinction. However, this
effect reverses longward of a transition wavelength $\lambda_t$.
The transition wavelength $\lambda_t\sim 0.34\ \micron$ for $50\%$
silicate/$50\%$ graphite, $\sim0.25\ \micron$ for $50\%$
silicate/$50\%$ AC1, and $\sim0.20\ \micron$ for $100\%$ silicate.
For these three examples, $\lambda_t\approx(2.1\pm0.6)\aeff$.

Although in this section we focused on clusters with $\aeff=0.127\micron$,
similar behavior is seen for other cluster sizes. The transition
wavelength at which the effect of porosity reverses increases when
the overall size of aggregate clusters increases.
Based on limited numerical experiments, we conjecture that, in general,
the transition wavelength occurs at $\lambda_t\approx(2.5\pm1.0)\aeff$:
porosity appears to increase the extinction
cross section per unit solid material for
$\lambda\lesssim1.5\aeff$, while {\it reducing} the extinction per unit solid
material at $\lambda\gtrsim3.5\aeff$.

It is interesting to note that for the
$50\%$ silicate/$50\%$ AC1 case, similar trends are seen in
\citet[][fig. 2]{Voshchinnikov+Ilin+Henning+Dubkova_2006}, where
the porous dust grain is spherical and modeled using the EMT-Mie
theory.
The fact that porosity actually decreases the extinction efficiency at
$\lambda > \lambda_t \approx 2\aeff$ was also seen by \citet{West_1991},
who used two types of aggregates with
different porosity and compared with the results of equal-volume
spheres, and found that porosity caused a decrease in extinction
efficiency when the equivalent size parameter is $2\pi R/\lambda
\le 5$ \citep[i.e., fig. 3 in][]{West_1991}.


\subsection{\label{subsec:monomer_size}
            Effects of monomer size/sensitivity to porosity}

In the above discussion, we fixed the monomer size to be $200$
\AA\ in radius. There is no direct evidence of how large the
monomers should be, hence we need to know if different monomer
size inside a ballistic aggregate will affect the overall
absorption and scattering cross sections.  To this end, we choose
two different monomer sizes, $100$ \AA\ and $400$ \AA, for the
BAM2 clusters with 2048 monomers and 32 monomers respectively.
These results
will
be compared with our fiducial BAM2 clusters
with 256 $a_0=200$ \AA\ monomers discussed in
\S\ref{subsec:dda}: we are comparing three clusters with the same $\aeff$,
i.e.,
composed of the same amount
of solid materials. We focus on the 50\% silicate/50\% graphite
case. For every cluster we average over three realizations, and
many random orientations of each realization. We used $\ndsph\sim
100$ for the 2048-monomer clusters and $\ndsph\sim 400$ for the
256-monomer and 32-monomer clusters. To satisfy the DDA accuracy
criterion equation (\ref{eq:dda_c1}) we only focus on results with
wavelength $\lambda\ge 0.2\ \micron$. We note that although the
three clusters contain identical volumes of solid materials with
$\aeff=0.127\ \micron$, the porosity is different for the 32, 256,
2048 clusters, with $\poro\approx 0.45$, $0.58$, and $0.64$,
respectively.

We show the results in Fig.\ \ref{fig:monomer_effect}, for $Q_{\rm
ext}$, $Q_{\rm abs}$, $Q_{\rm sca}$ and $g\equiv
\bracket{\cos\theta}$ respectively. Since the three clusters have
same amount of solid materials but different porosity (porosity
increases with increasing number of monomers), this behavior is
very similar to the general behavior of BA, BAM1 and BAM2 clusters
as shown in Fig. \ref{fig:DDA}. Once again, there is a
``transition'' wavelength
$\lambda_t\approx 3\aeff$
where the
calculated $Q_{\rm ext}$ is approximately independent of ${\cal
P}$ (the $\aeff=0.127\ \micron$ clusters in Fig.\ \ref{fig:DDA}
have $\lambda_t\approx0.34\ \micron\approx 2.7\aeff$, and the
$\aeff=0.127\ \micron$ clusters in Fig.\ \ref{fig:monomer_effect}
have
$\lambda_t\approx 0.42\ \micron\approx3.3\aeff$).

Although in Fig. \ref{fig:monomer_effect} there seems to be a
second transition wavelength at $\sim 1.5\ \micron$, we caution
that it might be artificial because the accuracy of our DDA
results decreases at the longest wavelengths (see Fig.
\ref{fig:converg}) for small $n_{\rm dip}$. This is particularly
true for the 2048 case where we used $n_{\rm dip}\approx 100$ due
to computational limits.

To further demonstrate that the effect seen in Fig.
\ref{fig:monomer_effect} is actually the result of varying the
porosity, we carry out another test. In Figure
\ref{fig:porosity_effect_Q2} we compare two clusters with
different geometries (BAM1 vs. BAM2), and different monomer sizes
(504\,\AA\ vs. 200\,\AA), but with the same amount of
material ($\aeff=0.160\micron$), and approximately the same porosity
($\poro\approx0.62$).
The two clusters have very
similar cross sections for extinction, absorption, and scattering.
Thus we conclude that the important parameters are just $\aeff$
(i.e., the amount of solid material) and the porosity $\poro$, and
monomer size does not have significant effects as long as the
monomer size $a_0\ltsim \lambda/2\pi$.

\subsection{Comparison with the MLS Approximation}\label{subsec:mls}

The DDA computations described above are very
time-consuming.
To model porous, composite dust grains, some authors have used the
analytical multi-layer sphere (MLS) model
\citep{Voshchinnikov+Mathis_1999,
       Voshchinnikov+Ilin+Henning_2005,
       Voshchinnikov+Ilin+Henning+Dubkova_2006},
where the sphere is composed of concentric spherical shells, each
of which is further composed of a set of spherical layers of
single composition. The number of such shells should be large
enough so that the results are unaffected by changing the order of
layers inside each
shell; in other words, the materials are well-mixed
inside such spheres. The problem of light scattering and
absorption by a multi-layer sphere can be solved
by a fast
algorithm \citep[e.g.,][]{Wu+Guo+Ren+etal_1997}.
The applicability of the MLS model to the aggregates considered here
has not
previously
been examined.

To compare the DDA results with the MLS model, one needs to know
the vacuum fraction $\fvac$, which is the volume fraction of
vacuum in the multi-layer sphere. Proponents of the MLS method
have not addressed the question of what value to use for $\fvac$
when modeling random aggregates; the optimal value of $\fvac$ is
not necessarily equal to the porosity $\poro$ given by eq.\
(\ref{eq:poro}). Therefore we test the MLS using different values
of $\fvac$ to find the value of $\fvac$ that minimizes the
difference between the DDA results and the MLS results. In
practice we set $\fvac=0.1 - 0.9$ with an increment of $\Delta
\fvac=0.1$ in the MLS calculations. We set the number of shells in
the MLS to be 40 so that the results do not vary significantly
when the order of layers inside each shell is changed.

We plot the results of MLS calculations on top of the DDA results
in Fig. \ref{fig:DDA} as black lines, which are bounded by the
black dashed line with $\fvac=0.1$ and the black dotted line with
$\fvac=0.9$. MLS results using intermediate vacuum fractions lie
between these two bounds. One can immediately see that the MLS
model is not a good approximation, especially for the 50\%
silicate/50\% graphite case. For the 50\% silicate/50\% AC1 case,
the MLS results have two transition wavelengths ($\sim0.31\,
\micron$ and $\sim0.95\,\micron$) at which the effect of porosity
reverses, as has already been reported by
\citet{Voshchinnikov+Ilin+Henning+Dubkova_2006}, although
$\aeff=0.1\,\micron$ in their work and $0.127\,\micron$ here.
However, this is different from our DDA results where there is
only one transition wavelength, near $\sim0.25\,\micron$. Similar
results are found for the $100\%$ silicate case. For the $50\%$
silicate/$50\%$ graphite case the MLS model fails to predict such
a transition altogether -- increasing the porosity increases the
MLS extinction at all wavelengths between 0.1 and $4\, \micron$.
For all three compositions, the MLS model seems to underestimate
the extinction at short wavelengths and overestimate the
extinction at long wavelengths. Our DDA results are of limited
accuracy (errors of up to $\sim8\%$) at long wavelengths for
$n_{\rm dip}\approx 400$ (see Fig. \ref{fig:converg}), but the
exact extinction values are even lower at these wavelengths, hence
the MLS errors are even larger than shown in Fig.\ \ref{fig:DDA}.

To quantify the deviation of the MLS results from the DDA results
we define a global error averaged over all 32 wavelengths:
\begin{equation}\label{eq:glob_err}
({\rm global\ error}\ Q_x)^2= \langle[\ln(Q_{x,{\rm
MLS}}/Q_{x,{\rm DDA}})]^2\rangle\ .
\end{equation}

Fig. \ref{fig:MLS_DDA_global_err} shows the global errors of
$Q_{\rm ext}$, $Q_{\rm abs}$, and $Q_{\rm sca}$ for the three
types of aggregates and three compositions, averaged over 5
realizations. Dotted lines show the standard deviation from the 5
realizations. It is clear that the accuracy of the MLS model
decreases when the porosity increases (from BAM2 to BA clusters)
or when the constituent materials become more absorptive (from
100\% silicate to 50\% silicate/50\% graphite).


\subsection{Comparison with the EMT Approximation}\label{subsec:emt}

One approach to estimate the optical properties of random
aggregates is to approximate them by homogeneous spheres, with an
``effective'' refractive index obtained from ``effective medium
theory'' (EMT); the scattering and absorption by the homogeneous
sphere is then calculated using Mie theory. Effective medium
theory comes in more than one variant; here we consider the form
of EMT developed by Bruggeman \citep[see][]{Bohren+Huffman_1983}
where the effective dielectric permittivity $\epsilon_{\rm eff}$
is calculated via
\begin{equation}\label{eq:bruggeman}
\sum_if_i\frac{\epsilon_i-\epsilon_{\rm
eff}}{\epsilon_i+2\epsilon_{\rm eff}}=0\ ,
\end{equation}
where $f_i$ and $\epsilon_i$ are the volume fraction and
dielectric permittivity of each composition, including vacuum.
Note that $\epsilon_{\rm eff}$ calculated in this way is not
affected by the detailed structure of the composite grain, i.e.,
``monomer size'' does not matter. If the aggregate contains $n$
distinct materials (including vacuum), then $\epsilon_{\rm eff}$
is a root of an $n^{th}$ order complex polynomial equation.  We
find that there is always just one root with ${\rm
Im}(\epsilon_{\rm eff}) \ge 0$ -- this is the physically
meaningful solution.

The cross sections calculated using EMT are compared to the DDA
results in Fig. \ref{fig:DDA_EMT} using our fiducial $N=256$
clusters with $a_0=0.02\ \mu$m, and the global errors from the EMT
approximation are shown in Fig. \ref{fig:EMT_DDA_global_err}. We
take the vacuum fraction in the EMT-Mie model to be $f_{\rm
vac}=0.1-0.9$ with $\Delta \fvac=0.05$. The EMT-Mie approach
provides much better agreements with the DDA results than the MLS
approximation does. This is particularly true for the
silicate-graphite and the silicate-AC1 compositions, where one
material is highly absorptive. It is clear from Figs.
\ref{fig:DDA} and \ref{fig:DDA_EMT} that the difference between
MLS and EMT-Mie increases when $\fvac$ increases, which is also
evident in fig. 2 of
\citet{Voshchinnikov+Ilin+Henning+Dubkova_2006} for the
silicate-AC1 case. Our ballistic aggregates have porosities ${\cal
P}\gtrsim 0.6$, hence it is not surprising that the MLS model is
not a good approximation.

The global error plots in Fig. \ref{fig:EMT_DDA_global_err} show
that the EMT-Mie model gives the optimal results
when\footnote{These are not the exact values of porosity at which
the EMT-Mie model fits the DDA results best, because our grid of
vacuum fraction has a coarse grid size of $\Delta f_{\rm
vac}=0.05$.} $\fvac=0.80,\ 0.70,\ 0.55$ for BA, BAM1, BAM2 clusters
respectively: it appears that for the three types of aggregates
the optimal vacuum fraction for EMT-Mie calculations is \beq
\label{eq:fvac=0.94poro} f_{\rm vac}\approx 0.94\poro\ . \eeq
The global errors of $Q_{\rm
ext}$ etc. are $\lesssim 10\%$ at this optimal $\fvac$, which
suggests that the EMT-Mie model provides a fairly good
approximation for computing overall cross sections. We plot the
difference between the EMT-Mie results (for optimal choice of
$\fvac$) and the DDA results, as functions of wavelength, in Fig.
\ref{fig:EMT_DDA_details}. It shows that although the EMT-Mie
results do not follow the DDA results exactly, the maximum
deviations are typically $\sim 20\%$ in all cases. It is also
apparent that the EMT-Mie model produces more forward scattering
at short wavelength, a property that is further discussed in Paper
II, which examines the angular distribution and polarization of
the scattered light.


\section{Summary and Discussions}\label{sec:discussion}
\begin{figure*}[h]
\begin{center}
\includegraphics[width=0.70\textwidth]{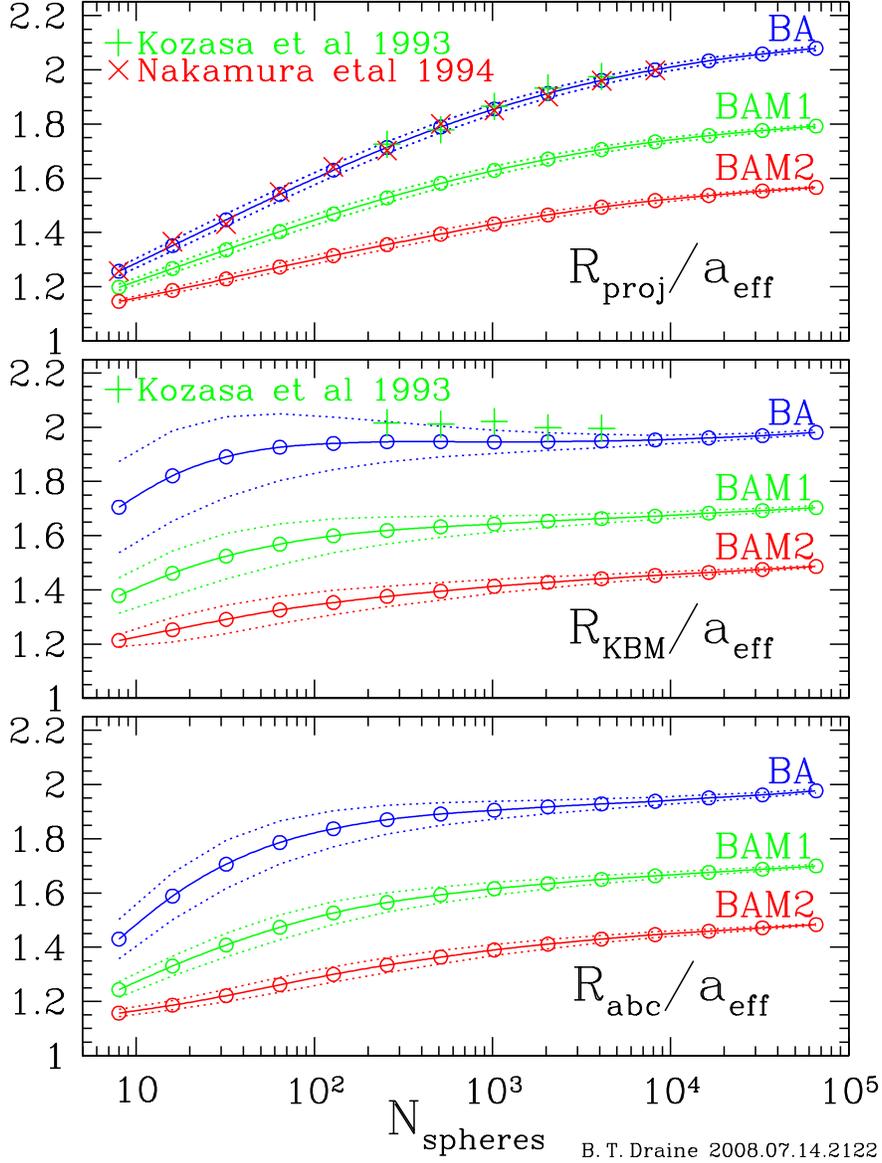}
\caption{\label{fig:frchar}
    Characteristic radius $\Rabc$, $\RKBM$, and the
    radius $\Rproj$ based on mean projected area,
    for clusters produced by BA, BAM1, and BAM2
    aggregation rules with single-size spheres (see text).
    Solid lines connect the mean values.
    Dotted lines show the $\pm1\sigma$ dispersion for random realizations.
    For BA clusters, we show
    $\RKBM$ and $\Rproj$ reported by
    \citet{Kozasa+Blum+Okamoto+Mukai_1993}, and $\Rproj$ given by
    \citet{Nakamura+Kitada+Mukai_1994}.
    }
\end{center}
\end{figure*}

\begin{figure*}[h]
\begin{center}
\includegraphics[angle=270,width=\textwidth]{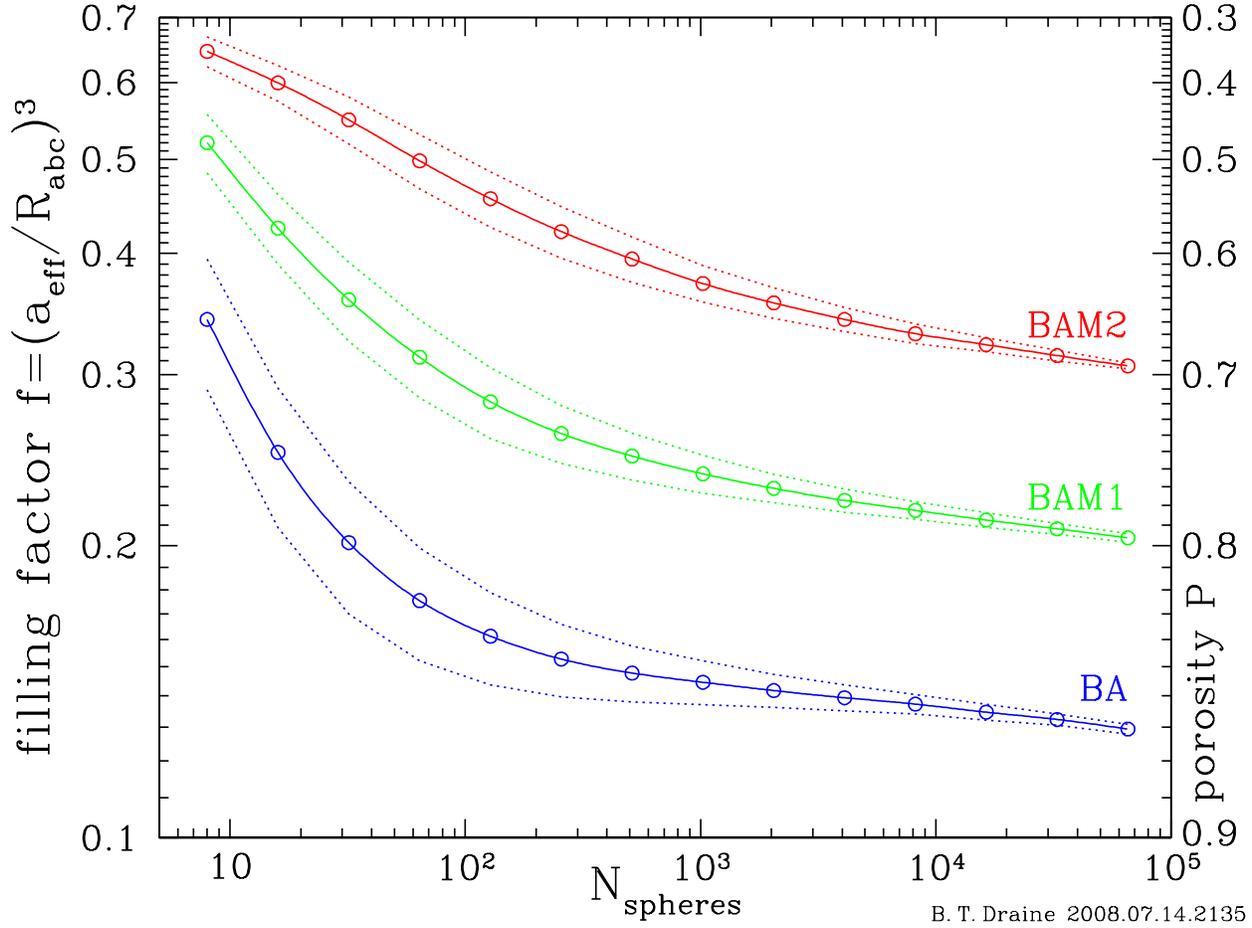}
\caption{\label{fig:ffill}
     Volume filling factor $\ffill$ (left scale) and
     porosity $\poro$ (right scale)
     for clusters produced by BA, BAM1, or BAM2
     aggregation rules with single-size spheres.
     Solid lines connect the mean values.
     Dotted lines show the $\pm1\sigma$ dispersion
     for random realizations.
     BA clusters are the least dense, with porosity $\poro\approx 0.87$ in the
     limit $N\rightarrow\infty$.
     BAM2 clusters are the most dense, with $\poro\approx 0.70$ in the
     limit $N\rightarrow\infty$.
     }
\end{center}
\end{figure*}

\begin{figure*}[h]
\begin{center}
\includegraphics[width=0.7\textwidth,angle=270]{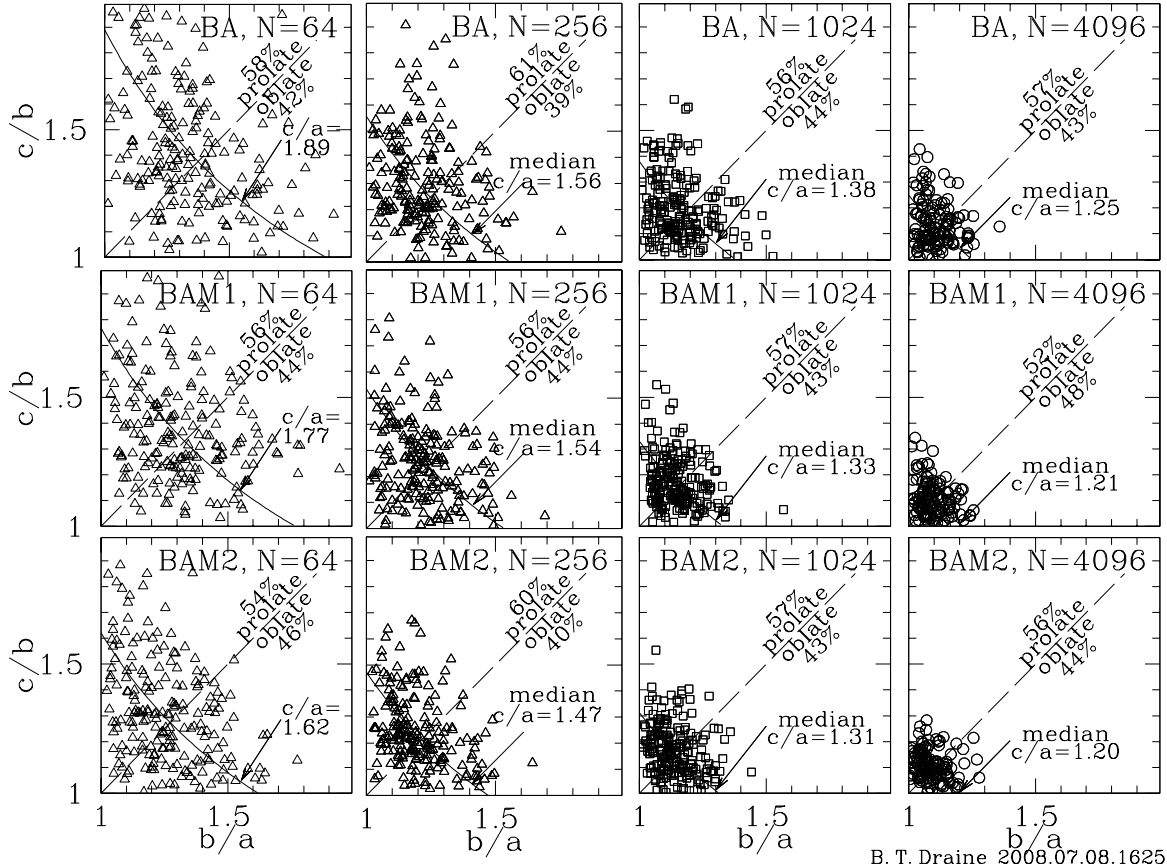}
\caption{\label{fig:axrat} Axial ratios for BA, BAM1, and BAM2
     clusters of $N$ spheres, for
     $N=64$, 256, 1024, and 4096.
     BA clusters are more asymmetric than BAM1 clusters, which in turn
     are more asymmetric than BAM2 clusters.
     Prolate shapes ($c/b > b/a$) appear to be slightly favored,
     and there is an overall tendency for clusters to be rounder
     as $N$ increases.
     }
\end{center}
\end{figure*}

\begin{figure*}
\centering
\includegraphics[width=\textwidth]{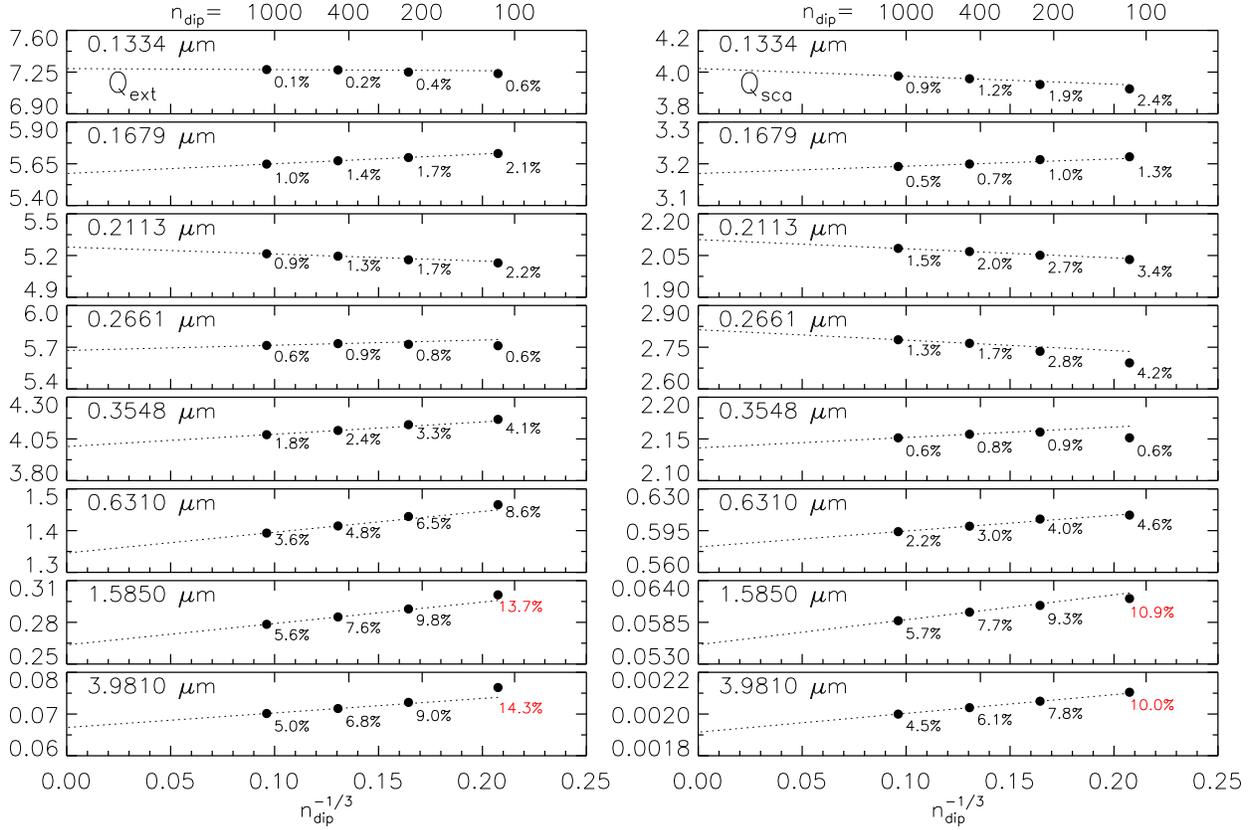}
\caption{\label{fig:converg}
   Convergence tests for the DDA method. Shown here are the
   extinction (left) and scattering (right) efficiencies for
   one $N=256$, $a_0=0.02\ \mu$m, $50\%$
   silicate and $50\%$ graphite
   cluster with $\poro=0.853$,
   computed using $n_{\rm dip}\thickapprox 100,\ 200,\ 400,\ 1000$,
   for selected wavelengths
   (as marked on top-left of each panel, in units of $\mu$m),
   and for a single orientation. The filled circles are the results,
   and the dotted lines are extrapolations to $\ndsph\rightarrow \infty$
   using the
   $\ndsph\thickapprox 400$ and $1000$ results (see text).
   The
   percentage alongside each data point is the fractional error with
   respect to the extrapolated value at $\ndsph\rightarrow
   \infty$.
   }
\end{figure*}

\begin{figure*}
\centering
\includegraphics[width=\textwidth]{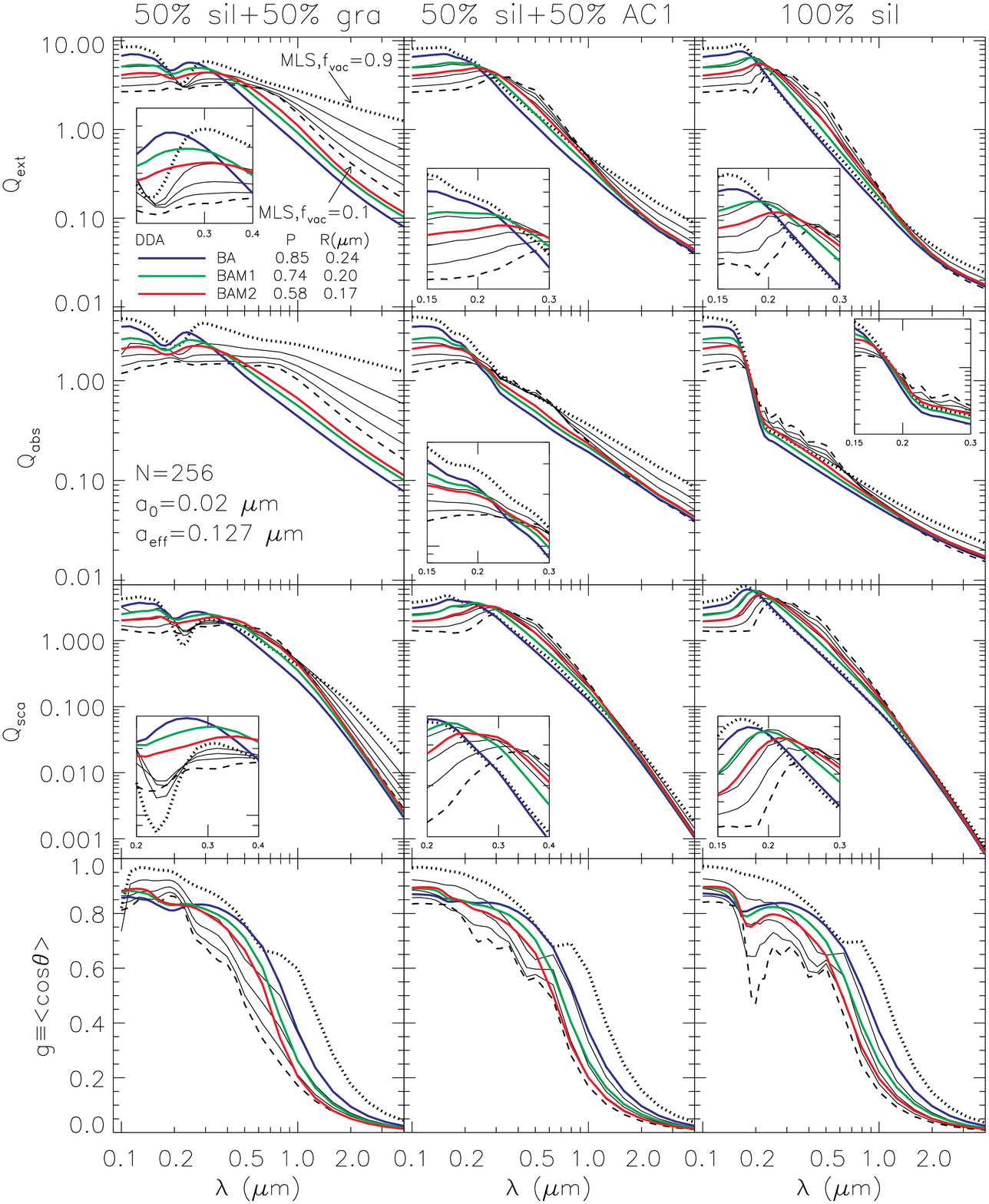}
\caption{\label{fig:DDA}
   Wavelength dependence of orientation-averaged $Q_{\rm ext}$,
   $Q_{\rm abs}$, $Q_{\rm sca}$ and $g\equiv\langle\cos\theta\rangle$ for three
   compositions and three aggregate types, for the fiducial $N=256$
   and $a_0=0.02\ \mu$m clusters with $\ndsph\approx400$.
   Each type of
   cluster
   is averaged
   over 5 realizations. Also plotted are the results of the MLS
   model: the black dashed lines are the MLS results
   with $\fvac=0.1$, the black dotted lines are for $\fvac=0.9$, and
   the black solid lines are for  $\fvac=$ 0.3, 0.5, 0.7.}
\end{figure*}

\begin{figure*}
\centering
\includegraphics[width=\textwidth]{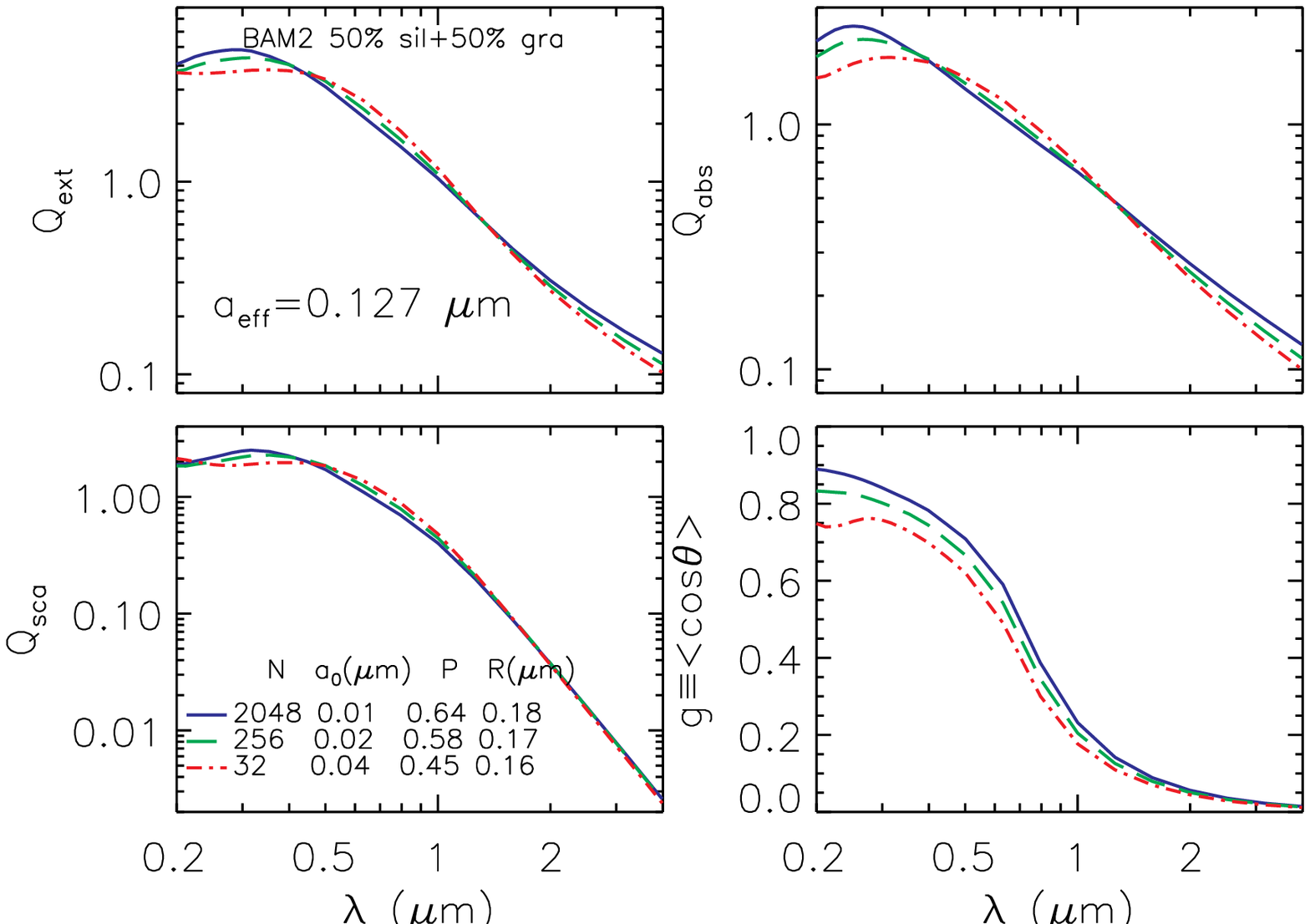} 
\caption{\label{fig:monomer_effect}
   $Q_{\rm ext}$, $Q_{\rm abs}$, $Q_{\rm sca}$ and
   $g\equiv \bracket{\cos\theta}$,
   for three BAM2 aggregates with $50\%$ silicate and $50\%$ graphite.
   These aggregates have the same amount of solid materials
   ($\aeff=0.127\micron$) but different porosities
   ($\poro\approx 0.45,\ 0.58,\ 0.64$) for $N=32,\ 256,\ 2048$.
   At a given $\lambda$, the computed $Q_{\rm abs}$, $Q_{\rm ext}$,
   and $Q_{\rm sca}$ are porosity-dependent, except at the
   ``transition radii'' $\lambda_t\approx 0.35\micron$ and $\sim1.4\micron$
   where the cross sections are insensitive to porosity.
   For $0.35 \ltsim \lambda \ltsim 1.4\micron$,
   the cross sections {\it decrease} as the porosity is increased.
   }
\end{figure*}

\begin{figure*}
\begin{center}
\includegraphics[width=\textwidth]{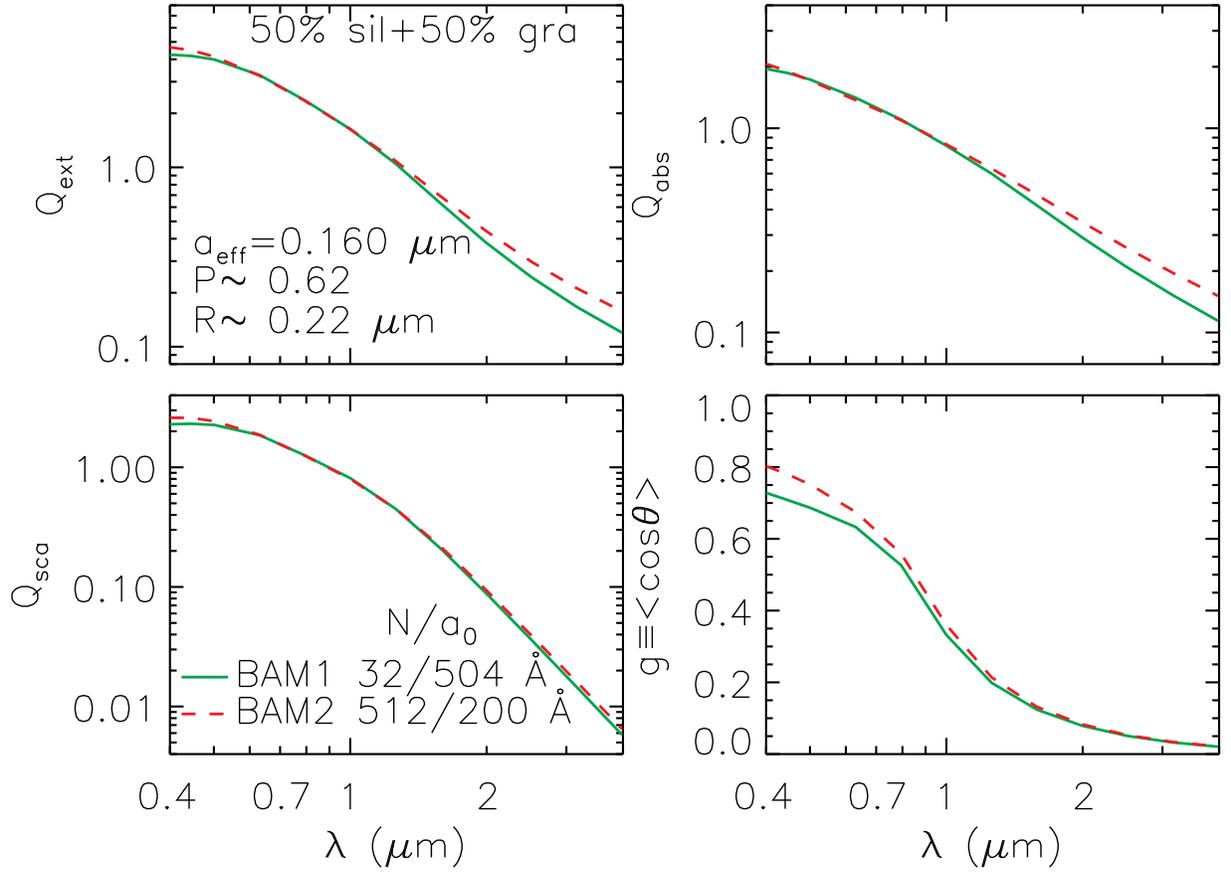}
\caption{\label{fig:porosity_effect_Q2}  Comparison of two
clusters with different geometries (BAM1 vs. BAM2), different
monomer size but similar porosity $\poro\approx0.62$ and same
$\aeff=0.160\ \micron$. The optical properties are very similar.
Therefore monomer size has no significant effects as long as
monomers are smaller than the incident wavelength.}
\end{center}
\end{figure*}

\begin{figure*}
\centering
\includegraphics[width=\textwidth]{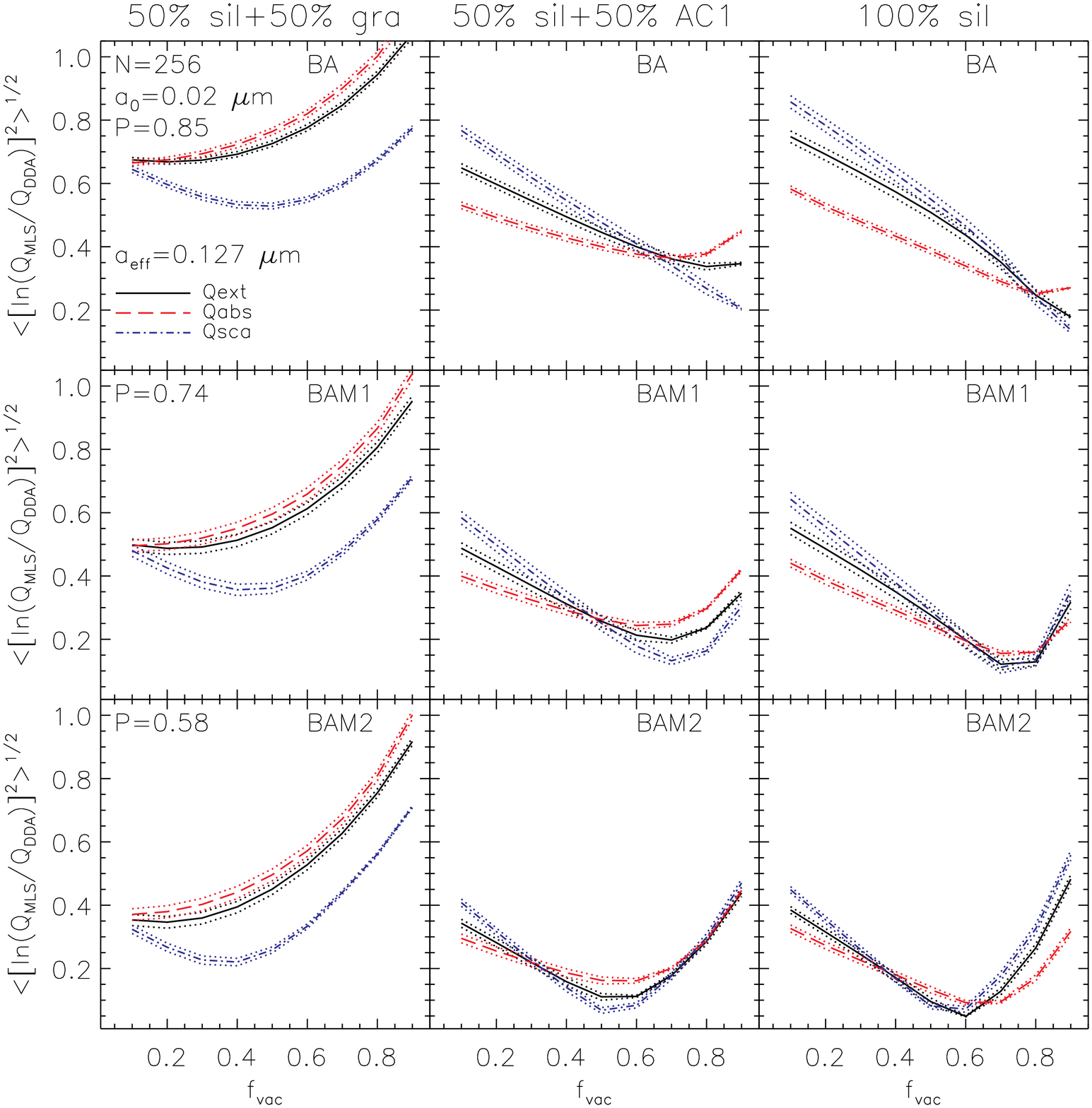}

\caption{Global errors (defined in eqn. \ref{eq:glob_err}) for the
MLS results and the DDA results for the fiducial aggregates as
shown in Fig. \ref{fig:DDA}. Dotted lines show the standard
deviation from the 5 realizations.} \label{fig:MLS_DDA_global_err}
\vspace*{0.7cm}
\end{figure*}

\begin{figure*}
\centering
\includegraphics[width=\textwidth]{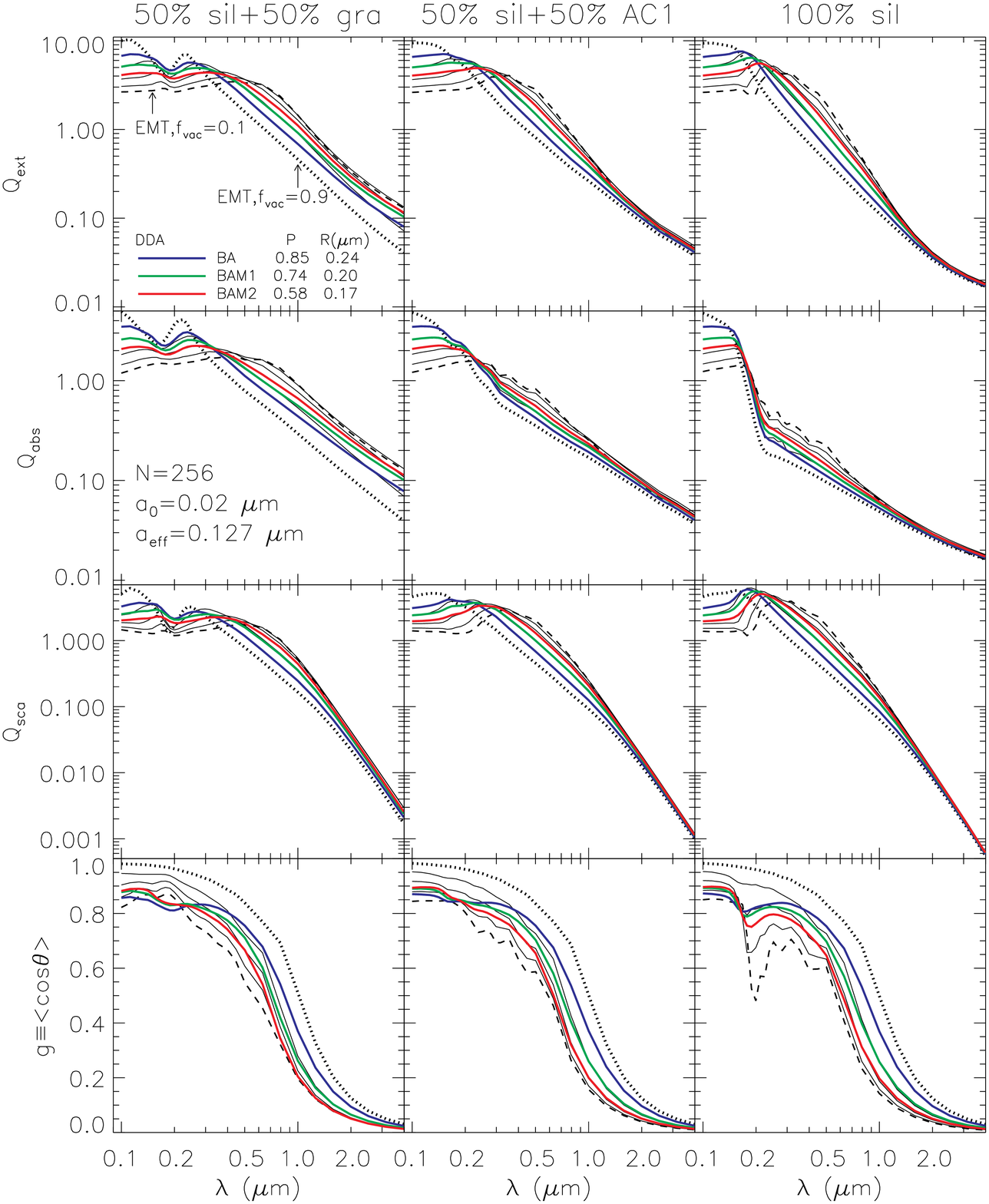}
\caption{The same as Fig. \ref{fig:DDA}, only with the MLS results
         replaced by the EMT-Mie
     results, for $f_{\rm vac}$ from 0.1 to 0.9.}
         \label{fig:DDA_EMT}
\end{figure*}

\begin{figure*}
\centering
\includegraphics[width=\textwidth]{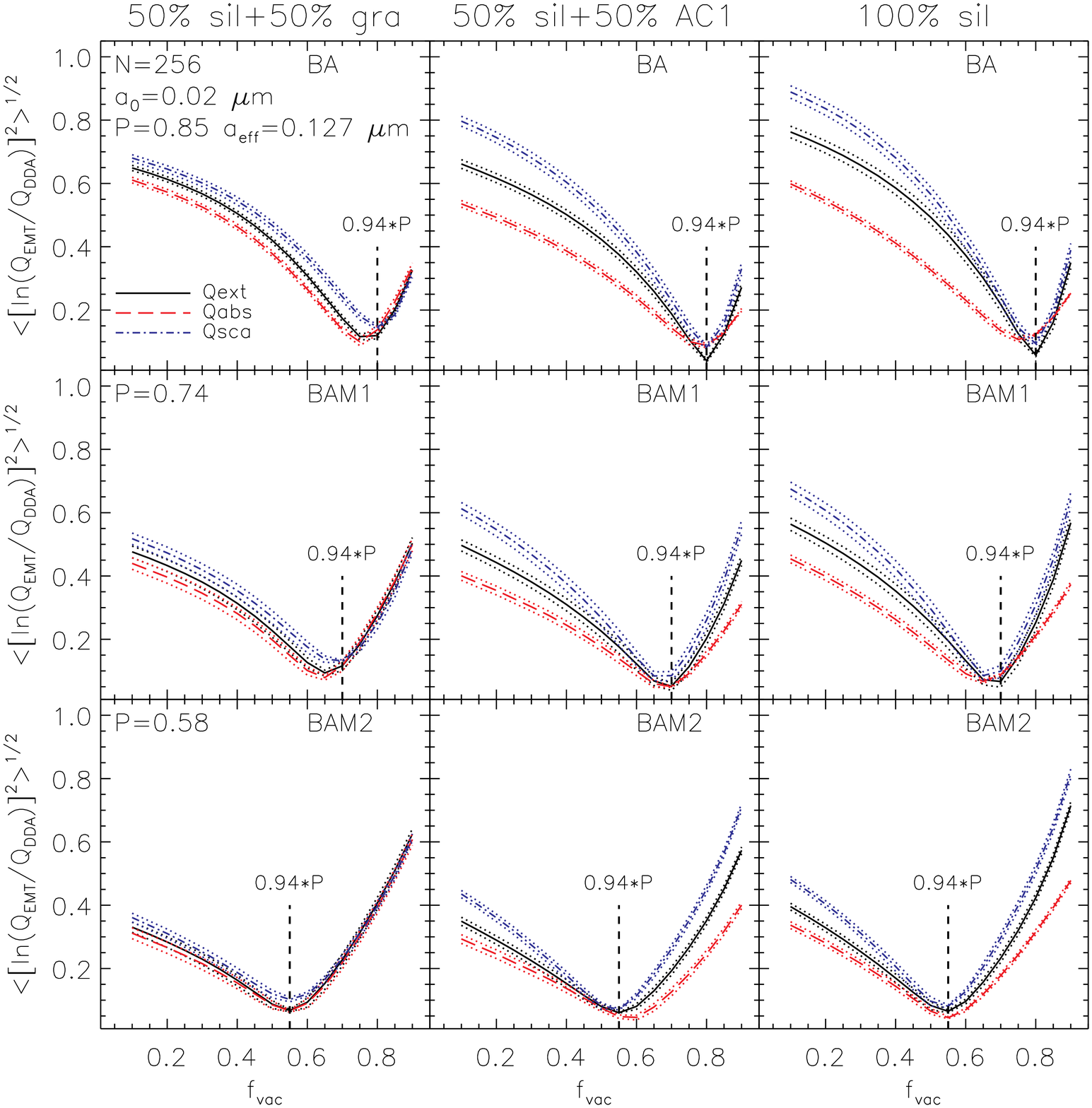}
\caption{Global errors (similar definition as in eqn.
    \ref{eq:glob_err}) for the EMT-Mie results and the DDA results for
    the fiducial aggregates as shown in Fig. \ref{fig:DDA_EMT}.
    $f_{\rm vac}$ from eq.\ (\ref{eq:fvac=0.94poro}) is shown.}
\label{fig:EMT_DDA_global_err}
\end{figure*}

\begin{figure*}
\centering
\includegraphics[width=\textwidth]{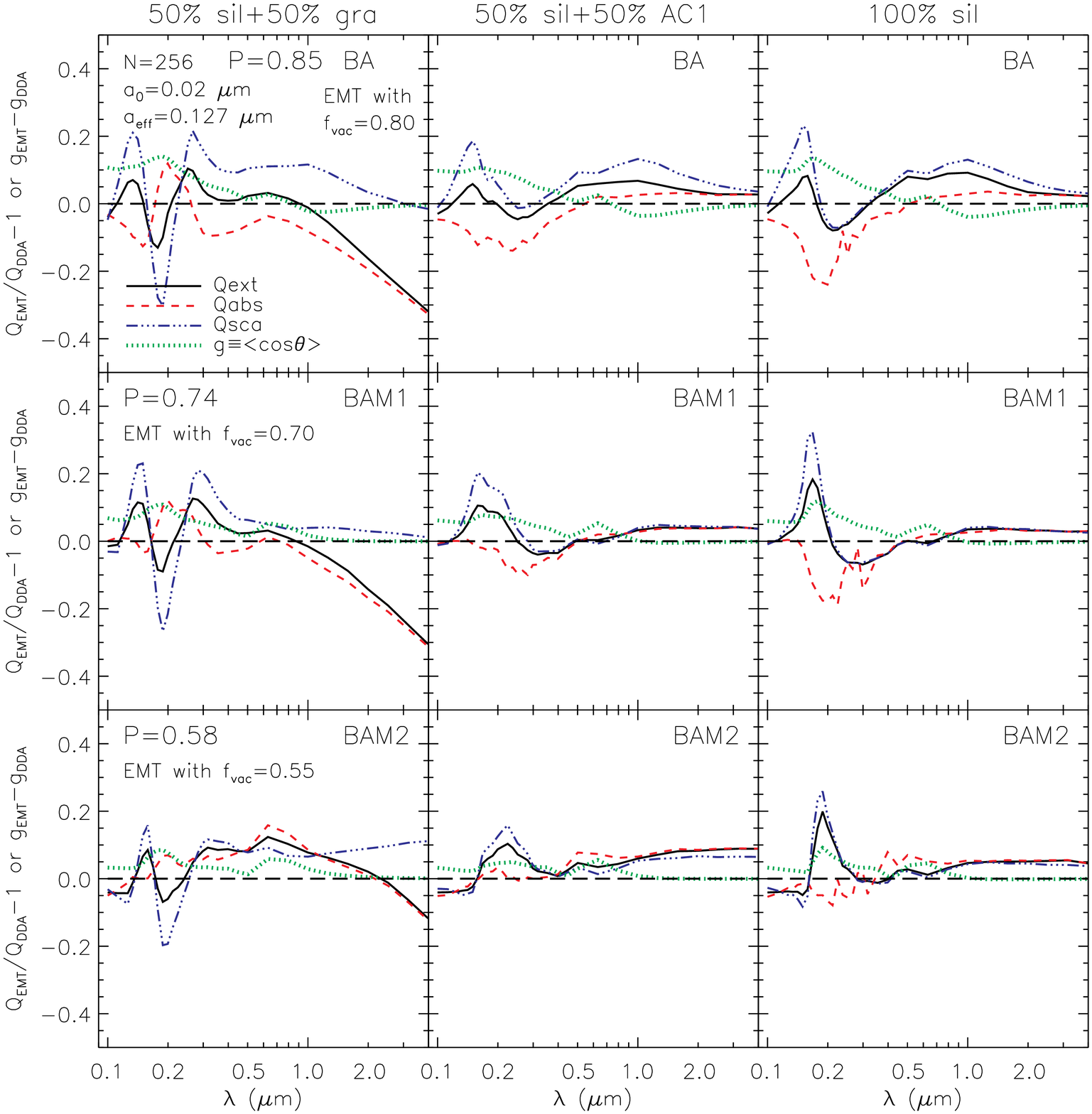} 
\caption{The difference between the optimal EMT-Mie results
    with $f_{\rm vac}=0.94\poro$ and
    the DDA results for our fiducial clusters. Plotted here are
    $(Q_{\rm EMT}/Q_{\rm DDA}-1)$ for cross sections, and $(g_{\rm
    EMT}-g_{\rm DDA})$ for $g\equiv \langle\cos\theta\rangle$. The
    fractional difference is typically
    $\lesssim 20\%$ for total cross sections.}
\label{fig:EMT_DDA_details}
\end{figure*}

The principal results of this study are the following:
\begin{enumerate}

\item Two new algorithms for generating random aggregates are
introduced: ballistic aggregation with one migration (BAM1), and
ballistic aggregation with two migrations (BAM2).  BAM1 and BAM2
aggregates are
less porous,
and more mechanically robust, than
conventional BA aggregates.

\item A measure $\poro$ of the porosity of a structure is proposed
(eq. \ref{eq:fill factor}), as well as a measure $\Rabc$ (eq.
\ref{eq:Rabc}) for the ``characteristic size'' of the structure.

\item Monte-Carlo simulations are used to determine the
statistical properties of $\poro$ and $\Rabc$ for ballistic
aggregates (see Tables 1-2 and Figs.\
\ref{fig:frchar}-\ref{fig:ffill}).

\item We confirm (see Figure \ref{fig:converg}) that the accuracy
of the DDA scales as the interdipole spacing $d$ as
$d\rightarrow0$, or, equivalently as $N_{\rm dip}^{-1/3}$ as
$N_{\rm dip}\rightarrow\infty$, where $N_{\rm dip}$ is the total
number of dipoles.

\item Scattering, absorption, and extinction cross sections are
calculated for the $N=256$, $a_0=0.02\ \micron$ ($\aeff=0.127\
\micron$) BA, BAM1, and BAM2 clusters, for three different
compositions: 100\% silicate, 50\% silicate/50\% amorphous carbon,
and 50\% silicate/50\% graphite, for wavelengths
$0.1\le\lambda\le3.981\,\micron$. The BA clusters (with the
highest porosity $\poro$) have the largest extinction cross
sections at short wavelengths, but at optical and near-IR
wavelengths, the BAM2 clusters (with the lowest $\poro$) provide
more extinction per unit solid material. At constant porosity and
same amount of solid material, the monomer size has no significant
effect provided the monomers are small compared to the incident
wavelength.

\item We compared the DDA results with the analytical MLS model
and EMT-Mie theory. We found the MLS model does not provide a good
approximation for absorptive and/or very porous grains; the
EMT-Mie model provides much better agreement with the DDA results.
For computing total cross sections ($Q_{\rm ext}$, $Q_{\rm abs}$,
$Q_{\rm sca}$), the EMT-Mie method provides results accurate to
$\sim10\%$ if the vacuum fraction $f_{\rm vac}$ is taken to be
$0.94\poro$.

\end{enumerate}

The effects of porosity on extinction cross sections have
important implications for the abundance budget problem in
interstellar dust models. The recent observed decrease of solar
abundances \citep[e.g.,][]{Asplund+Grevesse+Sauval_2005b} and the
claim that interstellar abundances might be better represented by
those of B stars \citep[e.g.,][]{Snow+Witt_1996} have imposed a
challenge to dust extinction models. Porous dust grains have been
thought to be a solution to this abundance budget problem
\citep[e.g.,][]{Mathis_1996}, because they were expected to result
in greater extinction per unit solid material than compact grains.
However, our results show that porosity actually {\it decreases}
the opacity at wavelengths long compared to the overall grain
size. Hence caution must be paid when dealing with the abundance
budget problem. Until detailed models have been constructed using
random aggregates to reproduce the observed interstellar
extinction (and polarization), we will not know if such grain
models will alleviate the interstellar abundance problem.  Work on
this problem is underway \citep{Johnson+Draine_2008}.

Ballistic aggregates are promising candidates for interstellar and
circumstellar dust grains. In the companion Paper II, we will
discuss the scattering properties of ballistic aggregates and
present examples that can reproduce the observations of light
scattered by dust in debris disks and comets.

\acknowledgements We thank the anonymous referee for helpful
comments. This research was supported in part by NSF grant
AST-0406883. Computations were performed on the Della and Artemis
computer clusters at Princeton University.

\bibliography{btdrefs}

\clearpage

\end{document}